\documentstyle[prb,aps]{revtex}

\tighten
\begin{document}
\draft
\widetext

\title{Vortex state in a $d$-wave superconductor}
\author{M. Franz, C. Kallin, P. I. Soininen}
\address{Department of Physics and Astronomy, McMaster University,
   Hamilton, Ontario, L8S 4M1 Canada}
\author{A. J. Berlinsky}
\address{Brockhouse Institute for Materials Science,  McMaster University,
   Hamilton, Ontario, L8S 4M1 Canada}
\author{A. L. Fetter}
\address{Department of Physics, Stanford University, Stanford, CA
94305}
\date{\today}
\maketitle

\begin{abstract}
We discuss the physics of the vortex state in a $d$-wave superconductor, using
the phenomenological Ginzburg-Landau theory, where
many novel phenomena arise from
the small admixture of the
$s$-wave component induced by spatial variations in the
dominant $d$-wave. Properties of an isolated vortex and of the Abrikosov vortex
lattice are studied by means of analytic and numerical methods. An
isolated vortex
has a considerable structure, with four ``extra'' nodes in the $s$-wave order
parameter symmerically placed around the core and an
amplitude forming a four-lobe
profile decaying as $1/r^2$ at large distances. The
supercurrent and magnetic field
distributions are also calculated. The Abrikosov lattice is in general oblique
with the precise shape determined
by the magnetic field and $s$-$d$ mixing parameter
$\epsilon_v$. The magnetic field distribution in the
Abrikosov state has two nonequivalent
saddle points resulting in the prediction of a double peak line shape in
$\mu$SR and NMR experiments as a test of a $d$-wave symmetry. Detailed
comparison
is made with existing experimental data and new experiments are proposed to
test for the predicted effects.
\end{abstract}



\section{Introduction}

After several years of debate there is growing agreement that the symmetry of
the order parameter in the high-$T_c$ cuprate superconductors is not a
conventional isotropic $s$-wave, but has a more complicated structure involving
nodes in the gap. Recent experiments sensitive to the phase of the
order parameter\cite{dexprev,dexp} provide
strong evidence for the $d_{x^2-y^2}$ symmetry with lines
of nodes along the $|k_x|=|k_y|$ directions. Support for the $d$-wave
symmetry also arises from specific heat measurements\cite{moler} and the recent
observation of a nonlinear Meisner effect.\cite{maeda} Photoemission studies,
\cite{photo} Josephson interference\cite{inter} and $c$-axis Josephson
tunneling\cite{caxis} experiments have been interpreted as being
inconsistent with
a pure $d$-wave order parameter. However most of these inconsistencies
 can be reconciled by allowing for states of mixed symmetry.\cite{bertouras}
In orthorhombic materials, such as YBCO and BiSCCO, if the dominant order
parameter is $d$-wave, a small $s$-component will be present
\cite{sigrist} even in a strictly uniform system. In tetragonal $d$-wave
materials, which will be considered in this work, this
$s$-wave component vanishes
identically in the bulk; however it may be nucleated locally by perturbations
which induce  spatial
variations of the $d$-wave order parameter,\cite{volovik,soininen}
 e.g., by external magnetic fields, surfaces or impurities.
In the present work we consider the vortex state of a $d$-wave superconductor
which results from applying a uniform magnetic field parallel to the $c$-axis
of the superconductor. We study the properties of isolated vortices and of the
Abrikosov vortex lattice, both of which which differ
in many aspects  from those found in conventional superconductors,
owing to the induced $s$-wave component.
These effects will play an important role in transport properties of
high-$T_c$ materials, which in turn are
crucial for all practical applications. Understanding
the static properties of $d$-wave vortices is a first important
step toward the description of the more complex dynamical effects in the
presence of transport currents, surfaces, impurities, {\em etc.}

The problem of an isolated vortex line in a $d$-wave superconductor was first
studied by Soininen, Kallin and Berlinsky\cite{soininen} who considered a
simple microscopic lattice model for electrons with on-site repulsion and
nearest neighbor attraction. The resulting Bogoliubov-de Gennes (BdG)
equations were solved numerically on finite clusters to obtain the order
parameter distribution for a single vortex. It was found that a substantial
$s$-wave component is nucleated near the vortex core with opposite winding
of phase relative to the $d$-component,\cite{volovik} and a distinct four-lobe
shape of the amplitude. These results were interpreted with help of the
phenomenological Ginzburg-Landau (GL) theory,\cite{joynt,kumar,doria}
where the non-zero
$s$ is driven by a mixed gradient coupling to the $d$-component. Ren, Xu and
Ting\cite{ren} later attempted a Gorkov-type derivation of the GL theory from
a continuum mean field model of  $d$-wave superconductivity and used the
resulting free energy to discuss the qualitative properties of a single
vortex. They obtained useful asymptotic expressions for the behavior
of the order parameters in various regions of the vortex.
Wang and MacDonald\cite{wang} investigated numerically
the electronic excitations
inside and outside  the cores of $s$-wave and $d$-wave vortices using
the self-consistent BdG equations. They found a distinctly different behavior
of the $T=0$ quasiparticle density of states in the core of the $d$-wave
vortex compared to that in the $s$-wave core within the same model.
 Very recently Ichioka
{\em et al.}\cite{ichioka} analyzed the structure of a $d$-wave vortex within
the    quasi-classical Eilenberger formalism. Their results appear to agree in
every aspect with the results of GL theory
presented below and in an earlier letter.\cite{berlinsky}
While the properties of an isolated vortex are now relatively
well understood, those of the vortex lattice have remained
largely unexplored.

Much of the work discussed above is based on a particular
(effective) microscopic model of superconductivity. However there
is presently no general agreement on the fundamental mechanism of pairing in
the high-$T_c$ cuprates.  A good alternative in such a situation is to study
the phenomenological GL theory, which is based only upon general concepts
related to symmetries of the system. Application of such theory to
conventional ($s$-wave) superconductors has demonstrated its ability
to predict virtually all of their phenomenological properties. In Section II,
we review the GL theory appropriate for the $d$-wave superconductor, which
involves both  $d$-wave  and an induced $s$-wave order parameter generated
through the mixed gradient coupling. We discuss some of the general properties
of this  free energy and  derive the corresponding GL differential equations
as well as an expression for the supercurrent. In doing this, and throughout
the entire paper, we restrict ourselves to the simple case of tetragonal
symmetry, described by the point group $D_4$.

Sections III and IV are devoted to the study of a single vortex  and of the
Abrikosov vortex lattice. Some of the results described here have
been previously
reported in a letter.\cite{berlinsky} Here we offer a more comprehensive
treatment of the problem, and we present a number of new results. For the
single
vortex we first review known analytical  results and complement these by
several
new observations. We then carry out numerical integration of the GL equations
for the single vortex geometry. In the region close to the vortex core our
results confirm previous work within the BdG framework.\cite{soininen} In
particular we find the induced $s$-wave order parameter which has the expected
four-lobe structure with minima along $\pm x$, $\pm y$ axes and maxima along
the $|x|=|y|$ diagonals, and the phase winding in the opposite sense relative
to the $d$-wave. Farther from the core the GL theory yields new results that
were inaccessible to the BdG treatment due to the cluster size limitations. At
a distance of several coherence lengths from the core the winding number
of the $s$-wave changes from $-1$ to $+3$ resulting in four ``extra'' nodes
in the $s$-wave order parameter
symmetrically placed along the $\pm x$, $\pm y$ axes.  Analysis of the
asymptotic solutions shows that these nodes are necessarily present in the
$s$-component, whenever pure $d$-wave solutions are thermodynamically stable
in the bulk. Our numerical work supports this conclusion. Quite generally the
distribution of the $d$-component, as well as the supercurrent and the magnetic
field, exhibit a four-fold anisotropy, the magnitude
 of which is proportional to the relative magnitude of $s$.

As was mentioned above, the problem of the vortex lattice, which forms in
magnetic fields close to the upper critical field $H_{c2}$, has not been
previously
addressed for a $d$-wave superconductor. In view of the four-fold anisotropy of
individual vortices one may expect that the conventional triangular Abrikosov
lattice\cite{abrikosov,kleiner} will be modified, especially since, even in the
absence of anisotropy, the difference in the free energy between  triangular
and square lattices is extremely small (less than 2\%). Moreover, recent
neutron
scattering\cite{keimer1} and STM\cite{renner} experiments reveal an oblique
vortex lattice in YBCO in strong magnetic fields. In Section IV we solve for
the
structure of the vortex lattice in the vicinity of $H_{c2}$. We generalize
the classic Abrikosov \cite{abrikosov} treatment to the $d$-wave case by first
minimizing the quadratic part of the free energy using a Gaussian variational
wave function, and then forming a periodic array of vortices from linear
combination of these functions. We include the effects of the vector potential
coupling self-consistently, thus improving upon our original calculation
\cite{berlinsky} which neglected these effects. The resulting vortex lattice
is found to be oblique, with an angle between primitive vectors ranging from
$60^\circ$ to $90^\circ$, depending on the strength of the mixed gradient
coupling and magnetic field and to a lesser extent on the other parameters
in the L free energy.

In Section V we summarize our results and discuss in some detail their
relevance to the existing experimental data. We also propose experiments that
might directly test some of our predictions.



\section{Ginzburg-Landau theory of a superconductor with
$\lowercase{d}$-wave pairing}

The Ginzburg-Landau (GL) theory for a superconductor with $d_{x^2-y^2}$
symmetry
has been  described by Joynt.\cite{joynt}
 The free energy density is expressed in terms of two components
of the order parameter, $s({\bf r})$ and $d({\bf r})$, with appropriate
symmetries, as follows
\begin{eqnarray}
f= \alpha_s|s|^2 &+& \alpha_d|d|^2 +\beta_1|s|^4 + \beta_2|d|^4 +
     \beta_3|s|^2|d|^2 +\beta_4(s^{*2}d^2 + d^{*2}s^2) \nonumber \\
  &+&\gamma_s|{\vec\Pi} s|^2 + \gamma_d|{\vec\Pi} d|^2
     +\gamma_v\bigl[ (\Pi_y s)^*(\Pi_y d) - (\Pi_x s)^*(\Pi_x d) + {\rm c.c.}
\bigr] +h^2/8\pi.
\label{fgl}
\end{eqnarray}
Here ${\vec\Pi}=-i\nabla -e^*{\bf A}/\hbar c$, and we assume that $d$ is a
critical order parameter, $i.e.$, we take $\alpha_s=\alpha' (T-T_s)$
and $\alpha_d =\alpha' (T-T_d)$ with
 $T_s<T_d$. The use of the same temperature derivative,
$\alpha'$, for $\alpha_s$ and $\alpha_d$ is justified below.
This also allows us to set $\alpha' = 1$ in the subsequent
analysis.
We assume that $\beta_1$, $\beta_2$, $\beta_3$, $\beta_4$,
$\gamma_s$, $\gamma_d$ and $\gamma_v$ are all positive
as it is suggested by a simple lattice model\cite{soininen} and a
Gorkov-type calculation within the continuum weak coupling theory.\cite{ren}
The parameters $\gamma$ are related to the effective masses in the usual way,
with $\gamma_i=\hbar^2/2m_i^*$, for $i=s,d,v$.
We shall be interested in the case when pure $d$-wave state is stable in
the bulk in the absence of perturbations, {\em i.e.}, situations when
$|d|>0$, $s=0$. The condition for such a state to be thermodynamically stable
is \cite{soininen}
\begin{equation}
\alpha_d<0,\quad 2\beta_2\alpha_s +(\beta_3-2|\beta_4|)|\alpha_d| >0.
\label{stability}
\end{equation}
With a finite $d$-component, the second transition temperature $T_s$ will be
renormalized by the fourth order invariants. In particular the transition to
the state with finite bulk $s$-wave component will occur at
\begin{equation}
T_s^*=T_s-{(\beta_3-2|\beta_4|)\over 2\beta_2-(\beta_3-2|\beta_4|)}
(T_d-T_s).
\end{equation}
Thus, even if the bare $T_s$ is close to $T_d$, the true transition
temperature $T_s^*$ may be much lower.
Moreover, when $ 2\beta_2-(\beta_3-2|\beta_4|)\le 0$, a
second transition will never occur and we may conclude that the precise value
of $T_s$ is not very important for the physics.

There are various ways of interpreting $f$, some of which have been
discussed by Joynt\cite{joynt} and by Volovik.\cite{volovik} Here
we provide an interpretation in terms of nearest neighbor bond
fields $v({\bf r})$ and $h({\bf r})$.
These fields describe the superconducting pairing amplitudes on the vertical
($v$) and horizontal ($h$) bonds of the square lattice representing the
crystalline structure of the cuprate superconductor, and  arise naturally
in the simple mean field lattice models of superconductivity with on site
repulsion and nearest neighbor attraction between electrons.
\cite{soininen,micnas} For tetragonal symmetry, the
free energy $f_b$ may be written in terms of these bond fields as
follows:
\begin{eqnarray}
f_b  = \alpha_0 \bigl( |v|^2 + |h|^2\bigr) &+& \epsilon \bigl( vh^* +
hv^*\bigr) + \gamma_L\bigl(|\Pi_xh|^2 + |\Pi_yv|^2\bigr)
     +\gamma_T\bigl(|\Pi_yh|^2 + |\Pi_xv|^2\bigr) \nonumber \\
 &+&  \gamma_C\left[(\Pi_xh)\bigl(\Pi_xv\bigr)^* +
                    (\Pi_yh)\bigl(\Pi_yv\bigr)^* + {\rm c.c.}\right]
  +   \beta(|h|^4 + |v|^4) +h^2/8\pi.
\label{fb}
\end{eqnarray}
In Eq.(\ref{fb}), $\alpha_0 = \alpha' (T-T_0)$, and $\epsilon$
stabilizes the relative phase of $v$ and $h$.  If $\epsilon$ is
positive, then a relative phase of $\pi$ is stabilized, and the
stable state has $d$-wave symmetry.  If $\epsilon < 0$, then
the quadratic terms in $f_b$ are minimized when $v$ and $h$ have the
same phase, giving a state with (extended) $s$ symmetry.  The
first two coefficients of the gradient terms $\gamma_L$ and
$\gamma_T$  involve derivatives along (e.g. $\Pi_yv$) and transverse
(e.g. $\Pi_xv$) to the bond directions.  In general, these two
coefficients will be different. The fourth order terms, proportional
to $\beta$, which are included in $f_b$ are the terms which would
arise in the mean field theory of XY spins.  In general, a mean
field theory for fermions will have other terms.  However it is
instructive to consider the consequences of these simple fourth
order terms, $i.e.$, $|h|^4 + |v|^4$.

The orthonormal transformation, $ s = (h+v)/\sqrt{2}$,
$ d = (h-v)/\sqrt{2}$, allows us to express the coefficients of
Eq.(\ref{fgl}) in terms of the coefficients in $f_b$.  The results
are:
\begin{equation}
\alpha_s = \alpha_0 - \epsilon, \qquad \alpha_d = \alpha_0 + \epsilon
\end{equation}
\begin{equation}
\beta_1=\beta_2=\beta_4 = \beta, \qquad \beta_3=4\beta
\end{equation}
\begin{equation}
\gamma_s=(\gamma_L+\gamma_T)/2 + \gamma_C, \quad
\gamma_d=(\gamma_L+\gamma_T)/2 - \gamma_C, \quad
\gamma_v=(\gamma_L-\gamma_T)/2
\label{gammas}
\end{equation}
The statement that the same value of $\alpha'$ occurs in
$\alpha_s$ and $\alpha_d$ is equivalent to the statement that the
temperature derivative of $\epsilon$ is negligible  in comparison to
the temperature derivative of $\alpha_0$.  If that is not the case,
then this approximation is not valid. In what follows we shall adopt the
above approximation for computational convenience, but we note that it is
in no way essential for the conclusions of this work, and  relaxing it only
leads to small quantitative changes.
The fourth order terms,
$|h|^4+|v|^4$, generate all of the terms in Eq.(\ref{fgl}) with
comparable magnitudes; in fact the resulting relative magnitudes of $\beta_i$'s
are very close to the weak coupling values.\cite{ren}
The mixed gradient term, $\gamma_v$, arises
from the {\it difference} in the coefficients of the longitudinal
and transverse gradient terms in the bond picture.  Of course, this
difference could be zero, but that is not expected on the basis of
symmetry.

To study the implications of the above  free energy (\ref{fgl}) for the
structure of the isolated vortex line and the vortex lattice
the first necessary step is to write down the field equations
for the order parameters. These are obtained in the standard
way by varying the
free energy (\ref{fgl}) with respect to conjugate fields $d^*$ and $s^*$.
We have
\begin{mathletters}
\label{eq:sd}
\begin{equation}
(\gamma_d \Pi^2 + \alpha_d) d + \gamma_v(\Pi_y^2-\Pi_x^2)s
+ 2\beta_2|d|^2d + \beta_3|s|^2d + 2\beta_4s^2d^* = 0,
\label{eq:d}
\end{equation}
\begin{equation}
(\gamma_s \Pi^2 + \alpha_s)s + \gamma_v(\Pi_y^2-\Pi_x^2)d
+2\beta_1|s^2|s + \beta_3|d|^2s + 2\beta_4d^2s^* = 0.
\label{eq:s}
\end{equation}
\end{mathletters}
In a similar manner, one obtains the current density in the $xy$ plane:
\begin{eqnarray}
{\bf j}=&&{e^*\hbar\over 2m_d^*}\bigl[d^*({\vec\Pi} d) +({\vec\Pi} d)^*d \bigr]
     +{e^*\hbar\over 2m_s^*}\bigl[s^*({\vec\Pi} s) +({\vec\Pi} s)^*s \bigr]
\nonumber \\
&-&{\hat x}{e^*\hbar\over 2m_v^*}\bigl[s^*(\Pi_x d) +(\Pi_x s)^*d +{\rm c.c.}
\bigr]
  +{\hat y}{e^*\hbar\over 2m_v^*}\bigl[s^*(\Pi_y d) +(\Pi_y s)^*d +{\rm c.c.}
\bigr].
\label{eq:j}
\end{eqnarray}
In carrying through the variational procedure it is necessary to
impose appropriate   boundary
conditions  for the superconductor-vacuum boundary. For  our two
component system these turn out to be
\begin{mathletters}
\label{bound}
\begin{equation}
{\bf n}\cdot[\gamma_d{\vec\Pi} d + \gamma_v(\hat y\Pi_y s -\hat x \Pi_x s)] =0,
\end{equation}
\begin{equation}
{\bf n}\cdot[\gamma_s{\vec\Pi} s + \gamma_v(\hat y\Pi_y d -\hat x \Pi_x d)] =0,
\end{equation}
\end{mathletters}
where ${\bf n}$ is the unit vector normal to the surface.
By combining the above two equations and comparing with the expression for the
current density (\ref{eq:j}), one can easily deduce that
\[
{\bf n}\cdot{\bf j}|_{ \rm boundary} =0,
\]
$i.e.$, the normal component of supercurrent vanishes,
as required on the  superconductor-vacuum boundary. We also note that for the
special case of a flat boundary along say the $yz$ plane, conditions
(\ref{bound}) acquire the simple form $\Pi_x s=\Pi_x d =0$, which is
analogous to the boundary condition in the usual one component system.

The above set of equations constitutes a complete Ginzburg-Landau theory for a
superconductor with $d_{x^2-y^2}$ pairing. This full theory is evidently too
complicated for most practical purposes, and one must resort to approximations
in order to obtain useful results. The rest of this paper is devoted to
two such approximations valid in  weak and strong magnetic fields.


\section{Near $H_{c1}$: isolated vortex line}


When the applied magnetic field $H$ is close to the lower critical field,
$H_{c1}$, spacing between individual vortex lines is large and it is
sufficient to consider structure of a single isolated vortex. As mentioned in
the Introduction, a single vortex line in a $d$-wave superconductor exhibits
rich and rather fascinating properties that have no analogue in conventional
superconductors with a single component order parameter. In the present section
we discuss these properties in some detail. First we review the analytical
results concerning the distribution of the order parameter, supercurrent and
magnetic field in various regions of the vortex. Second, we carry out an
explicit numerical integration of the GL equations for the single vortex
geometry to confirm and complement these analytic solutions.


\subsection{Analytic solutions}

As is appropriate in the case of high-$T_c$ cuprate
superconductors, we shall consider strongly type-II materials, in which
magnetic fields vary over length scale $\lambda$ that is much larger than the
relevant coherence length $\xi$ over which significant variations of the order
parameter can occur. In what follows we focus only on situations where
magnetic field is parallel to the $c$ axis of the superconductor.

For the problem of a single vortex line it will be convenient to work in the
cylindrical gauge expressed in the usual polar coordinates $(r,\varphi)$,
\begin{equation}
{\bf A}=\hat\varphi A(r),
\label{cylg1}
\end{equation}
with
\begin{equation}
A(r)={1\over r}\int_0^r r'h(r')dr'.
\label{cylg2}
\end{equation}
By adopting this particular gauge we restrict ourselves to magnetic fields
${\bf h}({\bf r})=\hat z h(r)$ that have no angular dependence. While this is
clearly
not exact for the $d$-wave vortex, we shall see that quite generally the part
of ${\bf h}$ that is not rotationally invariant is small and can thus be
computed as a perturbative correction to (\ref{cylg1}).

Let us first look at the behavior of the order parameter near the center of
the vortex, as $r\to 0$. In the relevant region where $r\ll\lambda$ the
magnetic
field can be treated as constant, $h\simeq h(0)\equiv h_0$ and the vector
potential becomes
\begin{equation}
A(r)={1\over 2} h_0r.
\label{vecpot1}
\end{equation}
For the singly quantized vortex $h_0$ can be roughly estimated by requiring
that the area $\sim \pi\lambda^2$ contains magnetic flux equal to a single flux
quantum $\Phi_0=hc/e^*$. This gives $h_0\simeq \Phi_0/\pi\lambda^2$.
The problem is now to find simultaneous solutions to the two GL equations
(\ref{eq:sd}) for $s$ and $d$, to leading order as $r\to 0$. Qualitatively
it is clear that at the core ($r=0$) both $d$ and $s$ vanish. Moving outward
from the core, the
amplitude of $d$ increases and generates nonzero $s$ via the mixed
gradient coupling. Around $r=\xi_d\equiv \sqrt{\gamma_d/|\alpha_d|}$ the
amplitude of $d$ starts to level off, attaining eventually its bulk value
$d_0\equiv \sqrt{|\alpha_d|/2\beta_2}$, which causes $|s|$ to
reach a maximum and
then decrease to 0 as $r\to\infty$. This qualitative picture suggests that in
order to study the leading order behavior we may first solve Eq.(\ref{eq:d})
for $d$ assuming  $s=0$, and then obtain the leading behavior of $s$ from
Eq.(\ref{eq:s}). With this assumption Eq.(\ref{eq:d}) becomes
\begin{equation}
(\alpha_d+\gamma_d\Pi^2)d + 2\beta_2|d|^2d =0,
\label{GLconv1}
\end{equation}
which is identical to the GL equation for the conventional one component
superconductor. The asymptotic solution to this equation near the core is well
known to be \cite{tinkham}
\begin{equation}
d(r,\varphi)\simeq (d_1r + d_3 r^3) e^{i\varphi},
\label{neard1}
\end{equation}
where constant $d_3$ is given by
\begin{equation}
d_3= -{d_1\over 8\xi_d^2}\left[1+2\pi\xi_d^2 h_0/\Phi_0\right],
\label{d3.1}
\end{equation}
and $d_1$ can be obtained by full integration of Eq.(\ref{GLconv1}).
Note that ordinarily only the leading dependence
$d(r,\varphi)\simeq d_1r e^{i\varphi}$ is quoted; however it turns out that
in our case the term $d_3r^3 e^{i\varphi}$ is necessary to obtain a consistent
expression for $s(r,\varphi)$. In Eq.(\ref{d3.1}) the factor
$\Phi_0/2\pi\xi_d^2$ dividing $h_0$ is of the order of the zero temperature
upper critical field $H_{c2}(0)$. Since we are interested in the region
close to $H_{c1}$, we have $h_0\ll H_{c2}$ and in what follows we shall
consistently  neglect terms $\sim h_0/ H_{c2}$ compared to unity. With this
simplification Eq.(\ref{d3.1}) becomes
\begin{equation}
d_3\simeq -d_1/8\xi_d^2.
\label{d3.2}
\end{equation}

The leading behavior of $s(r,\varphi)$ now can be obtained from the linearized
version of Eq.(\ref{eq:s}) which reads
\begin{equation}
(\alpha_s + \gamma_s\Pi^2)s + \gamma_v(\Pi_y^2-\Pi_x^2)d=0,
\label{GL2lin}
\end{equation}
by substituting for $d$ from Eq.(\ref{neard1}). Evaluating the action of the
$(\Pi_y^2-\Pi_x^2)$ operator in  polar coordinates gives
\begin{equation}
(\Pi_y^2-\Pi_x^2)d(r,\varphi) = -\left({e^*h_0\over\hbar c}\right)d_1r
e^{-i\varphi} + d_3r\left(3e^{-i\varphi} - e^{3i\varphi}\right),
\end{equation}
which suggests that the $s$-component is of the form
\begin{equation}
s(r,\varphi) = s_1 r e^{-i\varphi} + s_3r^3e^{3i\varphi}.
\label{nears1}
\end{equation}
Comparing the coefficients in front of different phase factors and again
neglecting terms $\sim h_0/H_{c2}(0)$, we obtain
\begin{equation}
s_1={3\over 8}\left({\gamma_v \over \alpha_s\xi_d^2}\right) d_1,
\label{s1}
\end{equation}
and, to the same order, $s_3=0$.

In summary, the leading order behavior of the order parameter near the core
is
\begin{mathletters}
\label{nearsd}
\begin{equation}
d(r,\varphi) = d_1 r e^{i\varphi},
\label{nearsd:d}
\end{equation}
\begin{equation}
s(r,\varphi) = {3\over 8} \left({\gamma_v\over \alpha_s\xi_d^2}\right)
d_1r e^{-i\varphi}.
\label{nearsd:s}
\end{equation}
\end{mathletters}
The most interesting feature of this result is the opposite winding of the
$s$-wave component relative to $d$. This was first pointed out by Volovik
\cite{volovik} based on a general symmetry argument. The solution of the form
(\ref{nearsd}) was also derived by Ren {\em et al.}\cite{ren}, however the
explicit form of the prefactor in the $s$-component is a new result of this
work. Knowledge of this prefactor will allow us to give a simple but accurate
estimate of the maximum $s$-wave amplitude, $s_{max}\equiv \max(|s|)$, induced
in the vicinity of the vortex core. In view of the fact that far outside the
core $s$ decays algebraically with $r$ (see below), such an estimate is quite
important for the assessment of the relative strength of the induced $s$-wave
component and the various new phenomena that its presence may lead to.
The estimate is based on the assumption that near the core $d$ and $s$ rise
over approximately the same length scale $\sim\xi_d$. In particular if we
assume that at $r=\xi_d$ the amplitude of $d$ is approximately half of its
bulk value \cite{tinkham} $d_0$, from Eq.(\ref{nearsd:d}) we have
$\xi_d d_1\simeq d_0/2$. Assuming further that $|s|$ attains its maximum also
around $r=\xi_d$ we arrive at the following estimate,
\begin{equation}
{s_{max}\over d_0} \simeq {3\over 16} {\gamma_v\over \alpha_s\xi_d^2}.
\label{smax}
\end{equation}
A similar estimate was given previously by us,\cite{berlinsky} based on a
simple argument involving the competition between the mixed gradient term and
other second order invariants in the free energy. This  argument gave the
correct
functional dependence on the GL parameters, however it missed  the numerical
prefactor $3/16=0.1875$, which is  important when investigating the
quantitative properties of the above solution. Comparison to the numerical
results (see the following subsection) shows that the above estimate
(\ref{smax}) is correct to
within about $15\%$, as long as $s_{max}< d_0/4$. When $s_{max}$ becomes
larger, the asymptotic solution (\ref{nearsd}) is no longer justified since
the condition $|s|\ll|d|$ is violated and our perturbative approach starts
to break down.

A noteworthy consequence of Eq.(\ref{smax}) is the temperature dependence
of $s_{max}$ near $T_d$. If we recall that close to $T_d$ we have
$d_0\sim\sqrt{1-T/T_d}$ and $\xi_d\sim 1/\sqrt{1-T/T_d}$, it follows that
\begin{equation}
s_{max}\sim (1-T/T_d)^{3/2}.
\label{smaxt}
\end{equation}
Faster decay of the $s$-component near $T_d$
compared to $d$ is a direct consequence of the fact that
as a non-critical order parameter the former is driven by the spatial
variations of the latter.\cite{comment1}
Thus, sufficiently close to $T_d$, the $s$-component will always be
negligible compared to $d$, and in many aspects a $d$-wave superconductor
will behave very much like a conventional single component superconductor.
Eq.(\ref{smaxt}) also self-consistently
justifies the above  perturbative solution of the GL
equations near the core which assumes $|s|$ to be small compared to $|d|$;
since $s_{max}/d_0\sim(1-T/T_d)$, the condition $|s|\ll|d|$ will be always
fulfilled sufficiently close to $T_d$.

The supercurrent and the local magnetic field near the vortex core can be
calculated from Eq.(\ref{eq:j}) using the order parameters given by
Eq.(\ref{nearsd}). We obtain, to  leading order in $r$,
\begin{eqnarray}
{\bf j}_s &=& d_1^2 {2\gamma_d e^* \over\hbar} \left[ 1-\left({3\over 8}
{\gamma_v\over \alpha_s\xi_d^2} \right)^2 \right] \hat\varphi r, \\
\label{nearj}
{\bf h}   &=& \hat z \left\{ h_0- d_1^2 {4\pi\gamma_d e^* \over c\hbar}
\left[ 1-\left({3\over 8}
{\gamma_v\over \alpha_s\xi_d^2} \right)^2 \right] r^2 \right\}.
\label{nearh}
\end{eqnarray}
Expression (\ref{nearj}) for the supercurrent shows explicitly that the
$s$-component with opposite winding of the phase relative to $d$ in
fact {\em diminishes} the total supercurrent, resulting in weaker shielding
of the external magnetic field compared to the conventional superconductor.

We next consider the region outside the core, $\xi_d\ll r\ll\lambda$.
We shall assume that in this region $d$ has already reached its limiting
form
\begin{equation}
d(r,\varphi)= d_0 e^{i\varphi}.
\label{dbulk}
\end{equation}
Because of the condition $r\ll\lambda$, the magnetic field can still be treated
to a reasonable approximation as constant, and the vector potential is thus
given by Eq.(\ref{vecpot1}). It is, however, easy to show that coupling to the
latter can be ignored in this region. In particular, rewriting all the relevant
operators in polar cylindrical coordinates, one can easily show that for
$d(r,\varphi)$ given by (\ref{dbulk}) it holds that
$\Pi^2 d(r,\varphi)= d_0 r^{-2} (-1 + r^2/\lambda^2)^2$. Clearly, the second
term in the brackets (which originates from the vector potential ${\bf A}$) can
be safely ignored with respect to unity, since, by assumption, $r/\lambda\ll
1$.
With some effort, one can  demonstrate that the vector potential is
also negligible  in the terms $(\Pi_y^2-\Pi_x^2)d(r,\varphi)$.

The problem of finding the asymptotic solution outside the core region reduces
to solving Eq.(\ref{eq:s}) for $s(r,\varphi)$ with $d$ given by (\ref{dbulk})
and ${\bf A}=0$, and the additional assumptions that  $|s|\ll|d|$ and
$|\nabla s|\ll
|\nabla d|$. These allow one to consider
only the linearized equation in which the relevant terms are
\begin{equation}
\gamma_v\left(\partial^2_x - \partial^2_y
\right) d + \alpha_s s + \beta_3 |d|^2 s + 2\beta_4 d^2 s^* =0.
\label{medlin}
\end{equation}
In polar coordinates,
\begin{equation}
\left(\partial^2_x - \partial^2_y\right)
d_0 e^{i\varphi} = {1\over 2r^2} \left(3 e^{3i\varphi} - e^{-i\varphi}\right)
d_0,
\label{mixf}
\end{equation}
suggesting that the solution to Eq.(\ref{medlin}) is of the form
\begin{equation}
s(r,\varphi)= {1\over r^2} \left(f_1  e^{-i\varphi} + f_3 e^{3i\varphi}\right).
\label{meds1}
\end{equation}
Substitution in Eq.(\ref{medlin}) then gives
\begin{eqnarray}
f_1 &=& {1\over 2} \gamma_v d_0 {(\alpha_s + \beta_3 d_0^2) + 6\beta_4 d_0^2
\over (\alpha_s + \beta_3 d_0^2)^2 - 4(\beta_4 d_0^2)^2 }, \label{f1} \\
f_3 &=& -{1\over 2} \gamma_v d_0 {3(\alpha_s + \beta_3 d_0^2) + 2\beta_4 d_0^2
\over (\alpha_s + \beta_3 d_0^2)^2 - 4(\beta_4 d_0^2)^2 }. \label{f2}
\end{eqnarray}
Asymptotic solution of this form was first obtained by Ren
{\em et al.}\cite{ren} From the knowledge of the order parameters
$d(r,\varphi)$ and $s(r,\varphi)$ one can compute the corresponding
distributions of the supercurrent and the magnetic field. In order to do this
consistently, one has to include corrections $\sim 1/r^2$ to the $d$-component
(such as were neglected in Eq.(\ref{dbulk})), as these are needed to  insure
that the continuity equation $\nabla\cdot{\bf j} = 0$
is satisfied. The resulting formulas can be found in Ref.\ [\onlinecite{ren}].

There are two important physical consequences of Eq.(\ref{meds1}). First, the
slow algebraic decay of the $s$-component outside the core region means that
asymptotically  in the presence of a vortex, the superconductor is not in a
pure
$d$-wave state, rather there is a small $s$-wave admixture with angle dependent
relative phase. As a result, fermionic excitations will be gaped in this
region. As demonstrated below, only at the length scale set by the penetration
 depth is the $s$-component cut off exponentially and a pure $d$-wave state is
established.

A second interesting property of the $s$-wave component can be obtained by
comparing the two solutions inside and outside the core. Inside the core
Eq.(\ref{nearsd:s}) implies that the winding number\cite{wind} of the
$s$-component is $-1$. The situation outside
the core is slightly more complex, but
near $T_d$ it holds that $f_3\simeq -3f_1$ [cf. Eqs.(\ref{f1}) and (\ref{f2})
in the limit $d_0\to 0$]. Thus the phase factor $e^{3i\varphi}$ in
Eq.(\ref{meds1}) will dominate
the behavior of $s(r,\varphi)$  and the winding number
far outside the core will be $+3$. For an analytic function the winding number
is a conserved topological quantity which can be changed only by the
presence of
a node. This forces us to conclude that four additional positive vortices
must exist outside the core in the $s$-component.\cite{berlinsky}
These ``extra'' vortices (or nodes) are a consequence of the topological
constraints imposed on the relative phases of $s$ and $d$ by the structure of
the GL equations (\ref{eq:sd}).
Behavior of the phases $\theta_s$ and $\theta_d$ is schematically depicted
in Fig.\ \ref{fig:ph1}, for the two asymptotic regions as given by Eqs.
(\ref{nearsd}), (\ref{dbulk}) and (\ref{meds1}). By inspection of the figure
one may conclude that the  four extra vortices are symmetrically placed on
$\pm x$ and $\pm y$ axes, since the $s$ component apparently changes
sign along these directions. Finally we
note that there are no extra nodes in the $d$-component and that
the total magnetic flux associated with one vortex line (consisting of 1
$d$-wave node and 5 $s$-wave nodes) is equal to one flux quantum; there is no
additional flux associated with the extra $s$-wave nodes.

We have argued above that the unusual nodal structure of the $d$-wave vortex
exists at temperatures close to $T_d$. It can be shown, however, that our
argument has much wider validity.  It is a simple matter to demonstrate that
a complex function of the form $g(\varphi)=ae^{-i\varphi}-be^{3i\varphi}$ with
$a$, $b>0$, will have winding number +3 for $b>a$ (and $-1$ for $b<a$).
Applying this criterion to $s(r,\varphi)$ given by Eq.(\ref{meds1}) and with
help of relations (\ref{f1}) and (\ref{f2}), one obtains the following
inequality
\begin{equation}
3(\alpha_s+\beta_3d_0^2) + 2\beta_4d_0^2 >
(\alpha_s+\beta_3d_0^2) + 6\beta_4d_0^2,
\end{equation}
as a requirement for the winding number $+3$ outside the core.
Upon expressing $d_0^2$ as $|\alpha_d|/2\beta_2$ and rearranging, one finds
that this inequality coincides with the stability condition (\ref{stability}).
It therefore follows that
 {\em for all} combinations of GL parameters consistent with stable
$d$-wave state, the asymptotic winding number of $s$ outside the core is
+3 and the non-trivial nodal structure described above exists.
We may conclude that the structure of the vortex core in a $d$-wave
superconductor is {\em inherently} much more complicated than that of a
conventional vortex. This statement is valid over the {\em entire} range
of temperatures in which the GL theory is applicable, in magnetic fields weak
enough  so that the vortex line can be considered isolated. Very recently,
the existence of the extra nodes has been confirmed by Ichioka {\em et al.}
\cite{ichioka}, who investigated the distribution of order parameters near
the vortex using the quasi-classical Eilenberger equations. On the other
hand, however, recent
numerical computations within the GL theory by Xu {\em et al.}\cite{xu} failed
to find evidence for this effect.
It would be most
interesting to see if evidence for the non-trivial vortex structure
can be established  in an experiment.

Finally we shall consider the region outside the core for $r\gg\lambda$.
In this region we may still assume the asymptotic form (\ref{dbulk}) for
$d(r,\varphi)$, but we can no longer treat the magnetic field as constant.
Taking into account the fact that $|s|\ll|d|$ in this region, we obtain the
usual London equation for the vector potential, which in the cylindrical gauge
(\ref{cylg1}) reads
\begin{equation}
\nabla^2{\bf A}= -{1\over\lambda^2}\left({\bf A}-{\Phi_0\over 2\pi r}
\hat\varphi\right).
\label{lond1}
\end{equation}
The asymptotic solution to this equation for $r\gg\lambda$ is
\begin{equation}
{\bf A}={\Phi_0\over 2\pi\lambda}\left[{\lambda\over r} -\left({\pi\over 2}
{\lambda\over r}\right)^{1/2} e^{-r/\lambda}\right]\hat\varphi,
\label{fara}
\end{equation}
which gives the usual exponentially decaying magnetic field far from the
vortex.\cite{tinkham} Using the vector potential given by (\ref{fara}) one can
solve for $s(r,\varphi)$ from Eq.(\ref{eq:s}). The result, to the leading
order in $(r/\lambda)$, is
\begin{equation}
s(r,\varphi)=\left({\pi\over 2}{\lambda\over r}\right)^{1/2} e^{-r/\lambda}
\left(s_1 e^{-i\varphi} + s_3 e^{3i\varphi}\right),
\label{fars}
\end{equation}
with
\begin{equation}
s_1=-s_3={1\over 2\lambda^2} {\gamma_v d_0 \over \alpha_s +(\beta_3-2\beta_4)
d_0^2}.
\label{s13}
\end{equation}
Thus, as expected, the $s$-wave component will be exponentially small beyond
distances from the core in excess of $\lambda$, and on these large length
scales the
$d$-wave superconductor will behave as a conventional single component
type-II material. Eqs.(\ref{fars}) and (\ref{s13}) also show that
to leading order, the total winding number of $s(r,\varphi)$ remains
undetermined (see the discussion of winding above). However, upon computing
higher order terms in $(r/\lambda)$ one finds that the winding number in this
region remains +3, so that there are no additional nodes present in the
$s$-component.


\subsection{Numerical results}

The analytic results presented in the preceding subsection establish rich and
complex structure of the vortex line in a $d$-wave superconductor; however
owing to the rather complex structure of the underlying GL equations
(\ref{eq:sd}) the analytic treatment is restricted to
limiting cases where certain small parameters can be identified. Consequently,
the information such a treatment provides is mainly of qualitative nature.
In order to study the problem in more detail, we have
integrated  the GL equations numerically. Besides confirming the above analytic
predictions, the numerical approach is capable of addressing the behavior of
the
order parameter at  length scales comparable to $\xi_d$, where the analytic
approach is difficult. In particular we will be most interested in the
detailed  behavior of the $s$-component near the core, focusing on its exotic
nodal structure that was predicted by topological arguments.

In order to arrive at a truly selfconsistent numerical solution, one should
in principle complement the GL Eqs.(\ref{eq:sd}) by the Maxwell equation
$\nabla\times{\bf h}_s={4\pi\over c}{\bf j}$ and include the induced magnetic
field ${\bf h}_s$ in the total vector potential ${\bf A}$. However, as we
are mostly
interested in the region near the core ($r\ll\lambda$), it is permissible
to neglect these screening effects and indeed the coupling to the vector
potential altogether, provided that we impose correct boundary conditions
for a single vortex geometry (see below). Neglecting the vector potential
leads to a significant
simplification of the problem. Physically this corresponds
to the extreme type-II limit, $\lambda/\xi\to \infty$.
 For a realistic system where $\lambda/\xi$ is
finite (but large), ignoring the vector potential coupling is equivalent
to neglecting terms $\sim(r/\lambda)^2$ compared to unity [see discussion
following Eq.(\ref{dbulk})].

With the vector potential absent from the GL equations it is convenient to
introduce a set of dimensionless quantities such that $\alpha_s$ is measured
in units of $|\alpha_d|$, $\beta$-parameters in units of $2\beta_2$, $s$ and
$d$ in units of the bulk $d$-wave gap $d_0$, and all the lengths in units
of $\xi_d$. This allows one to write the GL Eqs.(\ref{eq:sd}) in the following
simple dimensionless form
\begin{mathletters}
\label{gl:sd}
\begin{equation}
-\left(\nabla^2 +1\right) d + \epsilon_v(\partial_x^2-\partial_y^2)s
+ |d|^2d + \beta_3|s|^2d + 2\beta_4s^2d^* = 0,
\label{gl:d}
\end{equation}
\begin{equation}
-\left(\nabla^2 -\alpha_s\right) s + \epsilon_v(\partial_x^2-\partial_y^2)d
 +2\beta_1|s^2|s + \beta_3|d|^2s + 2\beta_4d^2s^* = 0,
\label{gl:s}
\end{equation}
\end{mathletters}
where $\epsilon_v=\gamma_v/\gamma_d$ and we have set $\gamma_s=\gamma_d$.
On physical
grounds [cf. Eq.(\ref{gammas})] we do not expect $\gamma_s$ and $\gamma_d$ to
differ dramatically; and we have verified that allowing $\gamma_s\neq\gamma_d$
does not have a significant effect on the solutions.

We have integrated Eqs.(\ref{gl:sd}) numerically on a rectangular $N\times N$
domain for the boundary conditions appropriate for a single vortex:
\begin{equation}
d|_{boundary}=d_0e^{i\varphi}, \ \ \ s|_{boundary}=0.
\label{bc1}
\end{equation}
We used an iterative Newton's algorithm as described in
Ref.\ [\onlinecite{newton}].
At each step of iteration the Conjugate Gradient Method\cite{golub}
was use to solve the resulting system of linear equations.

Results of our numerical analysis indeed confirm  all of the qualitative
features found by the analytic considerations of the preceding subsection.
Fig.\ \ref{fig:Samp} shows the behavior of the $d$- and $s$-wave amplitudes
near
the center of the vortex, with parameters described in the figure caption. The
resulting amplitude of the $s$-component for this particular parameter
configuration was $s_{max}=0.024d_0$, in good agreement with estimate
(\ref{smax}). A domain size of $N=201$ was used in the numerical integration,
encompassing a physical region of the size $L\simeq 20\xi_d$. In  Fig.\
\ref{fig:Samp} only the central ($121\times 121$) region is displayed, where
the
boundary effects are expected to be strongly suppressed (the numerical solution
was in fact well behaved all the way to the boundary of the system). As
expected
for this relatively weak admixture of the $s$-component, the amplitude of $d$
has almost perfect circular symmetry. The amplitude of $s$ is nearly circular
in the inner core of the vortex and it shows marked fourfold anisotropy
outside the core, in accordance with the asymptotic solutions (\ref{nearsd:s})
and (\ref{meds1}). Four symmetrically placed maxima along diagonals and four
nodes along $\pm x$ and $\pm y$ are visible in the contour plot. To see these
more clearly we show in Fig.\ \ref{fig:Scut} the amplitudes of the $s$-wave
component along $x$-axis and a $x=y$ diagonal. A node close to $x=3\xi_d$ is
clearly visible, which is nothing else than one of the four extra vortices.
The figure also confirms the linear behavior of $|s|$ and $|d|$ near the
origin and the fact that both rise on approximately same length scale $\xi_d$.
One can also see the $1/r^2$ decay of $|s|$ outside the core region, where
$|d|$ is constant.

Fig.\ \ref{fig:Sphsd} shows the superconducting phases $\theta_d$ and
$\theta_s$
of the two components of the order parameter. While the distribution of
$\theta_d$ looks very much like that of conventional singly quantized vortex,
the distribution of $\theta_s$ is clearly more complicated. In particular
the opposite winding of the phase near the core and four positive vortices
along the $\pm x$, $\pm y$ axes are clearly distinguishable. Comparison to
Fig.\
\ref{fig:ph1} shows that our numerical results are again in  complete
agreement with the analytical predictions summarized in the preceding
subsection.

The important quantity that determines the nature of excitations in the
vicinity of the vortex line is the relative phase
$\Delta\theta\equiv\theta_s-\theta_d$. We plot $\Delta\theta$ in Fig.\
\ref{fig:Sphrel}. Over much of the region the relative phase is $\Delta\theta=
\pm\pi/2$, resulting in a $d\pm is$ state that has minimum gap equal to
$|s|$. This is a direct consequence of the fact that $\beta_4>0$ in the
free energy (\ref{fgl}). However, the phase difference cannot be equal to
$\pm\pi/2$ over the entire area since this would be incompatible with the
topological constraints that require opposite winding of the two components
near the core. Thus narrow domain walls appear along the $\pm x$, $\pm y$
axes, where $\Delta\theta$ changes rapidly. This result is in agreement with
the microscopic treatment of Soininen {\em et al.}\cite{soininen} within the
Bogoliubov-de Gennes theory. However since the
complexity of this formalism did not
allow one to extend the calculations to sufficient distances from the core, the
extra vortices were originally not found. The present GL theory, being
inherently simpler allows us to study larger clusters. As one moves further
out from the core,  domain walls abruptly end at the cores of the four
$s$-wave vortices and $\Delta\theta$ starts to vary more slowly while being
still locked to $\pm\pi/2$ over large areas.

Supercurrent ${\bf j}$ produced by the above order parameter distribution,
computed
numerically from Eq.(\ref{eq:j}), is shown in Fig.\ \ref{fig:Sj}. Panel (a)
shows the distribution of the magnitude $|{\bf j}|$ while panel (b) displays
streamlines of the vector field ${\bf j}$. Note that because of the
Amp\`{e}re's
law $\nabla\times{\bf h}_s={4\pi\over c}{\bf j}$, the latter is equivalent
to the
lines of constant magnetic field given by the supercurrent, and thus Fig.\
\ref{fig:Sj}(b) also gives the distribution of the spatially varying component
of the screening field.

In addition to the particular case described above we have numerically
studied a
large number of other parameter combinations. All show similar behavior. The
feature that changes between different configurations is the relative
magnitude of the $s$-component, which is, as we have explicitly verified, well
described by Eq.(\ref{smax}). The larger the ratio $s_{max}/d_0$, the more
anisotropic $d$-wave component becomes and along with it the distribution of
supercurrent and induced magnetic field. As an example of such a case we show
amplitudes of $s$ and $d$ in Fig.\ \ref{fig:Samp2}, for the particular
parameter combination that yields $s_{max}\simeq 0.15d_0$. The relevant
supercurrent distributions is plotted in Fig.\ \ref{fig:Sj}.


\section{Near $H_{c2}$: structure of the vortex lattice}

In what follows we present our treatment of the vortex lattice problem.
In general we follow the path outlined by Abrikosov,\cite{abrikosov} with
necessary modifications that arise from the presence of two order parameters
in the free energy.


\subsection{Linearized GL equations and their variational solution}

In the vicinity of the upper critical field $H_{c2}$ the amplitudes of order
parameters are small, and the essential physics is contained in the linearized
field equations that are obtained from (\ref{eq:sd}) by neglecting the
nonlinear terms:
\begin{mathletters}
\label{eq:sdl}
\begin{equation}
(\gamma_d \Pi^2 + \alpha_d) d + \gamma_v(\Pi_y^2-\Pi_x^2)s = 0,
\label{eq:dl}
\end{equation}
\begin{equation}
(\gamma_s \Pi^2 + \alpha_s)s + \gamma_v(\Pi_y^2-\Pi_x^2)d  = 0.
\label{eq:sl}
\end{equation}
\end{mathletters}
Formally this corresponds to the expansion to leading order in the small
parameter $(H_{c2}-H)/H_{c2}$.
The gauge invariant gradient ${\vec\Pi}$ can be separated into two pieces,
\begin{equation}
{\vec\Pi}={\vec\Pi}_0 +{\vec\Pi}_s \equiv (-i\nabla-e^*{\bf A}_0/c\hbar) -
e^*{\bf A}_s/c\hbar,
\label{pivec}
\end{equation}
where ${\bf H}=\nabla\times{\bf A}_0$ corresponds to the uniform applied
field, and ${\bf h}_s=\nabla\times{\bf A}_s$ is the screening
 field induced by the supercurrent ${\bf j}_s$
in the sample, given by  the Maxwell equation
\begin{equation}
\nabla\times{\bf h}_s={4\pi\over c}{\bf j}_s.
\label{maxw}
\end{equation}
Let us for a moment ignore complications arising from the screening effects
and consider only the vector potential ${\bf A}_0$.
This is permissible, since as it will become clear later, corrections to
Eqs.(\ref{eq:sdl}) from the screening field are of the same higher order in
the small parameter $(H_{c2}-H)/H_{c2}$ as the nonlinear terms which have
been neglected in these equations. In the same spirit as the original Abrikosov
\cite{abrikosov} treatment, these higher order terms will be included
variationally in a later stage of the calculation.

It is easily seen that in the Landau gauge
${\bf A}_0={\hat y} Hx$ the linearized field equations (\ref{eq:sdl})
are satisfied by taking
\begin{equation}
d({\bf r})=e^{iky}d(x),\ \ s({\bf r})=e^{iky}s(x).
\label{dxsx}
\end{equation}
Thus, exactly as in the one component case first studied by Abrikosov,
\cite{abrikosov} we are left
 with one dimensional problem which can be stated as follows
\begin{mathletters}
\label{eq:sdlin}
\begin{equation}
\biggl\{\alpha_d + \biggl[{p^2 \over 2m} +{1\over 2}m\omega_c^2(x-x_k)^2\biggr]
\biggr\}d + \epsilon_v\biggl[ -{p^2\over 2m} +{1\over 2}m\omega_c^2(x-x_k)^2
\biggr]s = 0, \label{eq:sdlin:d}
\end{equation}
\begin{equation}
 \epsilon_v\biggl[ -{p^2\over 2m} +{1\over 2}m\omega_c^2(x-x_k)^2\biggr]d +
\biggl\{\alpha_s + \biggl[{p^2 \over 2m} +{1\over 2}m\omega_c^2(x-x_k)^2\biggr]
\biggr\}s = 0. \label{eq:sdlin:s}
\end{equation}
\end{mathletters}
Here we have denoted $p=-i\hbar\partial/\partial x$, $x_k=kl^2$, and
$\omega_c=(e^*H/mc)$.
The magnetic length $l=\sqrt{\hbar c/e^* H}$  will play the role of
characteristic length for the vortex lattice. We also assume henceforth
that $m^*_d=m^*_s \equiv m$, $i.e.,$ that
$\gamma_d=\gamma_s$, and we use $\epsilon_v = \gamma_v/\gamma_s = m^*_s/m^*_v$.
Equations (\ref{eq:sdlin}) resemble those for the quantum mechanical harmonic
oscillator problem
with the potential centered at $x=x_k$. In view of the fact that
$x_k$ is arbitrary, it is clear that Eqs.(\ref{eq:sdlin}) will have infinitely
many degenerate solutions which can be labeled by continuous index $k$.
This degeneracy will play a crucial role later when we construct the periodic
space-filling solution. However, for the moment, we shall ignore this issue and
investigate Eqs.(\ref{eq:sdlin}) with $x_k$ fixed.
The essential difference from the one component case is that these
equations have no obvious exact solutions. In what follows we shall seek
suitable variational solutions to Eqs.(\ref{eq:sdlin}).
In order to stress the analogy with the
harmonic  oscillator, we may write (\ref{eq:sdlin}) in the following way
\begin{mathletters}
\label{eq:sdlin2}
\begin{equation}
({\cal H}_0+\alpha_d) d + Vs = E d,
\label{eq:sdlin2:d}
\end{equation}
\begin{equation}
Vd + ({\cal H}_0+\alpha_s)s = E s,
\label{eq:sdlin2:s}
\end{equation}
\end{mathletters}
where ${\cal H}_0 = \hbar\omega_c(a^\dagger a +1/2)$ and
$V=\epsilon_v(\hbar\omega_c/2)
(a^\dagger a^\dagger +aa)$ are expressed in terms of the usual raising and
lowering operators, which can be
written as $a=[(x-x_k)/l + l(\partial/\partial x)]/\sqrt{2}$.  By including the
right hand side of Eqs.(\ref{eq:sdlin2})
we  are considering a slightly more general problem: $E=0$ corresponds to the
physical solution for $H=H_{c2}(T)$, and solutions for $E<0$ will be
useful later when we consider the
stability of various vortex lattice structures.

In order to motivate our variational solution to the linearized problem, let us
define
\begin{equation}
{\cal H}^\pm = {\cal H}_0 \pm V, \ \ \varphi^\pm = d \pm s.
\label{newar}
\end{equation}
In terms of these new variables, the set of equations (\ref{eq:sdlin2}) becomes
\begin{equation}
\left( \begin{array}{cc}
{\cal H}^++T-T^*  & -{\Delta T}/2 \\
                          &      \\
-{\Delta T}/2              & {\cal H}^-+T-T^*
\end{array} \right)
\left( \begin{array}{c}
{\varphi^+} \\ \\ {\varphi^-}
\end{array} \right)
=E \left( \begin{array}{c}
{\varphi^+} \\  \\ {\varphi^-}
\end{array} \right),
\label{eq:sdlin3}
\end{equation}
where we have defined
\begin{equation}
T^*=(T_d+T_s)/2, \ \ {\Delta T}=T_d-T_s,
\label{ts}
\end{equation}
for convenience in later calculations.
A nice feature of the representation (\ref{eq:sdlin3})
is that for the degenerate case ${\Delta T}=0$ the equations for
$\varphi^+$ and $\varphi^-$ decouple, each becoming a simple harmonic
oscillator problem.  Motivated by this fact
we shall seek the variational solution for the general case
in the form of normalized ground state
wave-functions of the harmonic oscillator,
\begin{equation}
\varphi_k^\pm(x)= \sqrt{\sigma_\pm \over l\sqrt{\pi}}
e^{-\sigma_\pm^2 (x-x_k)^2/2l^2}.
\label{varsol}
\end{equation}
The variational parameters $\sigma_+$ and $\sigma_-$  will be determined
by minimization of the eigenvalue\cite{footnote}
\begin{equation}
\langle E\rangle
= (T-T^*)  +{1\over 2}\langle{\varphi^+}{\cal H}^+{\varphi^+}\rangle
           +{1\over 2}\langle{\varphi^-}{\cal H}^-{\varphi^-}\rangle
           -{1\over 2}{\Delta T}\langle{\varphi^+}{\varphi^-}\rangle,
\label{Emin1}
\end{equation}
where angular brackets stand for spatial averages. All the integrals are
easily evaluated and if one defines
$\sigma_+=\sigma\cos\vartheta,\ \ \sigma_-=\sigma\sin\vartheta$,
the resulting expression for $\langle E\rangle$
 can be explicitly minimized with respect to
$\sigma^2$. The minimum occurs for
$\sigma^2=\tan\vartheta +{1/\tan\vartheta}$, and is
\begin{equation}
{\langle E\rangle \over{\Delta T}} = {T - T^*\over{\Delta T}}
+{1\over 4}\biggl({\hbar\omega_c\over{\Delta T}}\biggr)
\biggl[(1+\epsilon_v)\tan\vartheta +(1-\epsilon_v){1\over\tan\vartheta}\biggr]
- {1\over 2}\sqrt{{2\tan\vartheta \over 1+\tan^2\vartheta}}.
\label{Emin3}
\end{equation}
The last equation must be minimized numerically with respect to
$\tan\vartheta$. It is also clear from this equation that two parameters,
$\epsilon_v$ and
\begin{equation}
\Lambda=\hbar\omega_c/{\Delta T},
\label{lambda}
\end{equation}
determine the nature of the variational solution.
In the two limiting cases the exact minimum can be easily found.
In the low field limit,
$\Lambda\ll 1$, we have $\sigma_+\approx\sigma_-\approx 1$,
while in the high field limit, $\Lambda\gg 1$, we have $\sigma_\pm\approx
[(1\pm{\epsilon_v})/(1\mp{\epsilon_v})]^{1/4}$. It follows that at least
intermediate values of $\Lambda$ are required for appreciable effects from
$s$-$d$ mixing to occur. Otherwise  ${\varphi^+}\simeq{\varphi^-}$ and
according to  Eq.(\ref{newar}) the $s$-component effectively vanishes.

Solutions to Eq.(\ref{Emin3}) with $\langle E\rangle=0$ give the
dependence of the upper critical field $H_{c2}$ on the temperature.
Whenever a finite admixture of the $s$-component is present,
a characteristic upward curvature is found near the critical temperature in
$H_{c2}(T)$.
Such curvature has been observed experimentally in both
LSCO and YBCO compounds \cite{hidaka} and has been interpreted
as a consequence of $s$-$d$ mixing.\cite{joynt,doria} For given parameters
$T_d$ and $T_s$ and several values of $\epsilon_v$ such dependence is shown in
Fig.\ \ref{fig:hc2}, as obtained by numerical minimization of Eq.(\ref{Emin3}).


\subsection{ Vortex lattice solution}

To construct a periodic
vortex lattice consider a linear superpositions of the basis
functions (\ref{varsol}) of the form
\begin{equation}
\Psi_\pm({\bf r})=\sum_n c_n e^{inqy}\varphi_n^\pm(x),
\label{psipm}
\end{equation}
where $c_n$ are complex constants. In order to impose periodicity in
$y$-direction we have constrained the values of $k$ to integer multiples
\begin{equation}
k_n=nq, \ \ n=0,\pm 1,\pm 2,\dots
\label{kn}
\end{equation}
of the parameter $q$ which will be determined later from the requirement of
minimum free energy.  The space filling solutions of GL equations can
be written as, cf.\ Eq.(\ref{newar}),
\begin{eqnarray}
d({\bf r})&=& [\Psi_+({\bf r}) + \Psi_-({\bf r})]/2, \nonumber \\
s({\bf r})&=& [\Psi_+({\bf r}) - \Psi_-({\bf r})]/2.
\label{drsr}
\end{eqnarray}
These solutions
will also be periodic in $x$ provided that the constants $c_n$ satisfy
the condition $c_{n+N}=c_n$ for some integer $N$.
As  was first noted by Abrikosov,\cite{abrikosov} the analysis of the vortex
lattice for general $N$ is extremely difficult. It was however conjectured
that the
absolute minimum of the free energy takes place for $N\leq 2$, in which case
the analysis is simplified. In what follows we shall restrict ourselves to the
case of $N=2$ for the two component system.
Taking $N=2$ we have $c_{2n}=c_0$ and $c_{2n+1}=c_1$. This, along with
Eqs.(\ref{kn}) and (\ref{drsr}),  implies periodicity
of $s$ and $d$ in $x$ and $y$ with periods
\begin{equation}
L_x=2l^2 q, \ \ L_y= 2\pi/ q.
\label{periods}
\end{equation}
Each rectangular $L_x\times L_y$ unit  cell  then contains an amount of flux
\begin{equation}
HL_x L_y= 2(hc/e^*)\equiv 2\Phi_0
\label{flux}
\end{equation}
where $\Phi_0$ stands for the flux quantum. Thus, by construction, each
rectangular
unit cell contains exactly two singly quantized vortices, independent of the
value of parameter $q$. The resulting vortex
lattice may be thought of as centered
rectangular with two quanta per unit cell or, equivalently, as an oblique cell
with lattice vectors of equal length and one flux quantum.
While the restriction to centered rectangular lattices is made
primarily for the computational convenience, it is also compatible with
recent experiments on YBCO which show evidence\cite{keimer1,renner}
for oblique vortex lattices with nearly equal lattice constants in high fields.

The parameter $q$ controls the shape of the unit cell.
It is customary to define the ratio
\begin{equation}
R={L_x/ L_y} = (l^2/\pi)q^2,
\label{r}
\end{equation}
and it follows that $R=1$ corresponds to the square, $R=\sqrt{3}$
corresponds to the triangular, and the intermediate values $1<R<\sqrt{3}$
imply  the oblique  vortex lattice.

The solution that we have constructed for the GL equations (\ref{eq:sd})
has three free parameters, $c_0$, $c_1$ and $R$. These parameters determine
the structure of the vortex lattice near $H_{c2}$. Within the linearized
approximation to the GL free energy these solutions are degenerate in energy.
It is the fourth order terms that lift this degeneracy and determine the
equilibrium lattice structure.
 In order to find this  minimum
one must take into account the fourth order terms in the free energy
(\ref{fgl}) as well as the effects of screening which were so far ignored.

The complete average free energy density (\ref{fgl}) can be written as
\begin{equation}
\langle f \rangle = \langle f_2 \rangle + \langle f_4 \rangle +
\langle h^2 \rangle/8\pi,
\label{fgl24}
\end{equation}
where $f_2$ and $f_4$ stand for  quadratic and quartic invariants
respectively, and ${\bf h}={\bf H}+{\bf h}_s$ is the local magnetic field.
Let us now consider the effect of screening by looking at the gradient terms
in $\langle f_2\rangle$ with the complete ${\vec\Pi}$ as given by
Eq.(\ref{pivec}). A typical term will be of the form
\begin{equation}
\langle |{\vec\Pi} d|^2 \rangle = \langle |{\vec\Pi}_0 d+{\vec\Pi}_s d|^2
\rangle
\simeq \langle |{\vec\Pi}_0 d|^2 \rangle +
\langle {\vec\Pi}_s\cdot [d^*{\vec\Pi}_0 d + {\rm c.c.}] \rangle,
\label{expa}
\end{equation}
where in the last equality terms of the order of $|{\vec\Pi}_s d|^2$ have been
neglected. This is consistent with the general idea of GL theory of
keeping only  terms up to fourth order in the order parameters.
Being proportional to the supercurrent, ${\vec\Pi}_s$ already
contains terms quadratic in the order parameters. If we expand all
the remaining
gradient terms  in the similar way, systematically
neglecting terms containing order parameters to powers higher than four,
we can  write the result  as
\begin{equation}
\langle f_2\rangle = \langle f_2^{(0)}\rangle +{\hbar\over e^*}
\langle {\vec\Pi}_s\cdot{\bf j}_s\rangle.
\label{f2p}
\end{equation}
Here $ f_2^{(0)}$ is the part of $f_2$ containing only the
${\vec\Pi}_0$ piece of the gauge invariant gradient, $i.e.$, the
quadratic part in the absence of screening,  and similarly
${\bf j}_s$ is assumed to be given by Eq.(\ref{eq:j}) with
${\vec\Pi}={\vec\Pi}_0$.
If we take into account the property of the variational solution
$ \langle f_2^{(0)}\rangle =E \langle |s|^2+|d|^2\rangle$ that
follows from Eqs.(\ref{eq:sdlin2})  and use the definition of ${\vec\Pi}_s$
we can write
\begin{equation}
\langle f \rangle =  E \langle |s|^2+|d|^2\rangle -
(1/c)\langle {\bf A}_s\cdot{\bf j}_s\rangle
+ \langle f_4 \rangle + \langle ({\bf H}+{\bf h}_s)^2 \rangle/8\pi,
\label{fgl24a}
\end{equation}
The second term on the RHS can be simplified by expressing ${\bf j}_s$ through
 the Maxwell
equation (\ref{maxw}). Integrating by parts and neglecting the surface term one
obtains
$(1/c)\langle {\bf A}_s\cdot{\bf j}_s\rangle=\langle h_s^2\rangle/4\pi$.

Similarly the last term on the RHS can be rewritten recalling the definition
${\bf B}={\bf H}+\langle {\bf h}_s\rangle$ of the  magnetic induction  as
$ \langle h_s^2\rangle/8\pi-H^2/8\pi +{\bf B}\cdot{\bf H}/4\pi$.

The manipulations performed
above are useful since in  fixed applied magnetic field the proper
thermodynamic potential to minimize is the mean Gibbs free energy density
related to $f$
by $\langle g \rangle = \langle f \rangle -{\bf B}\cdot{\bf H}/4\pi$. For this
quantity we finally arrive at  an expression
\begin{equation}
\langle g \rangle =  E \langle |s|^2+|d|^2\rangle
+ \langle f_4 \rangle - \langle h_s^2 \rangle/8\pi-H^2/8\pi.
\label{ggl}
\end{equation}
Before we proceed with minimization of the Gibbs potential let us notice
that the simple thermodynamic relation $\partial\langle g \rangle/ \partial H
=-B/4\pi$ can be used to extract the average screening field
 in the superconductor
\begin{equation}
\langle h_s\rangle \equiv B-H =
 - \left({\partial E \over \partial H}\right) \langle |s|^2+|d|^2\rangle.
\label{ind}
\end{equation}
A similar relation between the average induced field  and the
order parameter  for the conventional
$s$-wave superconductor \cite{abrikosov}  is known as the ``first Abrikosov
identity,'' but the corresponding determination of the spatial distribution
of ${\bf h}_s({\bf r})$ is more complicated (see below).
 It is easy to verify that in the limit $\epsilon_v\to 0$ ($i.e.$ in
the limit of pure $d$-wave) Eq.(\ref{ind}) assumes the precise form of this
identity, including all the relevant prefactors that follow upon expressing
$\partial E/ \partial H$ from Eq.(\ref{Emin3}).
In the vicinity of $H_{c2}$ it holds that
$E\simeq(\partial E/\partial H)(H-H_{c2})$ and
it follows that to the leading order we can write
$\langle f_2^{(0)}\rangle =\langle h_s\rangle (H_{c2}-H)/4\pi$.
This allows us to  express the Gibbs free
energy in the form where the leading dependence on the magnetic field $H$ is
manifestly displayed:
\begin{equation}
\langle g \rangle-\langle g \rangle_n =
{1\over 4\pi}(H_{c2}-H)\langle h_s\rangle
+ \langle f_4 \rangle - {1\over 8\pi}\langle h_s^2 \rangle,
\label{ggl2}
\end{equation}
with $\langle g \rangle_n =-H^2/8\pi$ being the normal state contribution
 to the Gibbs free energy.

Consider now a simple scaling transformation $(s,d)\to (\lambda s,\lambda d)$
where $\lambda$ is a real number. It is clear that under such a transformation
$\langle h_s\rangle \to \lambda^2\langle h_s\rangle$, while
$\langle f_4\rangle \to \lambda^4\langle f_4\rangle$ and
$\langle h_s^2\rangle \to \lambda^4\langle h_s^2\rangle$. Consequently, the
Gibbs free energy (\ref{ggl2}) will have a well defined minimum for the
particular value of $\lambda$. We use this property
 to determine the normalization
of the order parameters $s$ and $d$. Carrying out the minimization we obtain
\begin{equation}
\langle g \rangle-\langle g \rangle_n =
-{1\over 8\pi}{(H_{c2}-H)^2\langle h_s\rangle^2 \over
8\pi \langle f_4 \rangle - \langle h_s^2 \rangle},
\label{ggl3}
\end{equation}
an expression which is clearly independent of the particular normalization
of $s$ and $d$.
If we further define the Abrikosov ratio $\beta_A$ and the Ginzburg-Landau
parameter $\kappa$ by
\begin{equation}
\beta_A={\langle h_s^2\rangle \over \langle h_s\rangle^2}, \ \ \ \
\kappa^2=4\pi{\langle f_4\rangle \over \langle h_s^2\rangle},
\label{betaA}
\end{equation}
we can write the resulting Gibbs free energy for the Abrikosov vortex lattice
in the familiar form \cite{abrikosov}
\begin{equation}
\langle g \rangle-\langle g \rangle_n =
-{1\over 8\pi}{(H_{c2}-H)^2 \over
(2\kappa^2-1)\beta_A}.
\label{ggl4}
\end{equation}
Several remarks are in order. The Abrikosov ratio $\beta_A$ defined
by Eq.(\ref{betaA}) is independent of the
coefficients $\beta_i$ in the quartic part of the free energy and depends only
on the shape of the unit cell in the vortex lattice. To the extent that
$\kappa$ is independent of the specific lattice shape, the minimum Gibbs free
energy corresponds to the minimum of $\beta_A$, which generalizes the
familiar Abrikosov result, apart from writing it in terms of magnetic field
instead of the absolute squared order parameter. As will be shown below
by numerical calculation, it is indeed true that the parameter $\kappa$ defined
by Eq.(\ref{betaA}) depends  only very weakly on the vortex lattice shape, and
thus the factor $(2\kappa^2 -1)$ in the denominator of Eq.(\ref{ggl4}) serves
simply as the criterion for  type-II behavior, which occurs only for
$\kappa > 1/\sqrt{2}$. It is in this sense that one can think of $\kappa$ as
a generalization of the conventional Ginzburg-Landau parameter;
we note that $\kappa$
 defined by Eq.(\ref{betaA}) cannot be simply related to the usual ratio of
penetration depth $\lambda$ to coherence length $\xi$.
This difficulty is related to the fact
that in the
presence of two order parameters $s$ and $d$ we have, strictly speaking,
two distinct coherence lengths, $\xi_s$ and $\xi_d$.
Most observable phenomena will only
reveal a single ``effective'' coherence length given by a certain combination
of $\xi_s$ and $\xi_d$, but this will presumably depend on the type of
probe used in the  experiment. By contrast, there will be only single
penetration depth $\lambda$, as this quantity is related to the decay of
magnetic field inside the superconductor. Alternatively, $\lambda$
 may be viewed as a measure of the bulk superfluid density, which is in
the  present case associated with the
$d$-wave component only, since the $s$-wave
vanishes in the bulk. Thus it may be suggested that $\kappa=
\lambda/\xi_A$, where $\xi_A$ is the effective coherence length relevant to the
Abrikosov lattice, determined by the usual criterion of overlapping vortex
cores at $H=H_{c2}$.


\subsection{Magnetic field distribution}

The ultimate goal of this section is to
determine the actual structure of the vortex lattice by minimizing the
Gibbs free energy given by  Eq.(\ref{ggl4}). To obtain the  parameters
$\beta_A$
and $\kappa$ that enter this expression it will be necessary to evaluate
the spatial averages  $\langle f_4\rangle$ and $\langle h_s^2\rangle$ (note
that quantity $\langle h_s\rangle$ has been already calculated in
Eq.(\ref{ind}) by thermodynamic argument). The former of the two averages
can be computed in a fairly straightforward manner since $f_4$ is directly
related to the vortex lattice solutions $\Psi_\pm({\bf r})$, which are simple
linear superpositions of the Gaussian wavefunctions $\varphi^\pm_k$ given by
Eq.(\ref{varsol}). The situation with the other average, $\langle
h_s^2\rangle$,
 is more complicated as one has to first
invert the Maxwell equation (\ref{maxw}) in order
to express the local screening field ${\bf h}_s({\bf r})$ in terms of the
supercurrent
${\bf j}_s$. Both of these quantities are  themselves of interest,
as they can be in principle  measured by various experimental probes
(see Sec.\ VI for the discussion).

With this in mind let us calculate the spatial distribution of
the screening field. If we express ${\bf h}_s$
in terms of the vector potential ${\bf A}_s$, the Maxwell equation (\ref{maxw})
can be written as
\begin{equation}
\nabla^2{\bf A}_s = -{4\pi\over c}{\bf j}_s,
\label{maxw2}
\end{equation}
where we have taken  advantage of the fact that  the Landau gauge satisfies
$\nabla\cdot{\bf A}_s=0$. The easiest way to invert the equation (\ref{maxw2})
is to exploit the periodicity of the vortex lattice solution
and work in  Fourier space. If we write
\begin{equation}
{\bf j}_s({\bf r})=\sum_{\bf k} e^{i{\bf k}\cdot{\bf r}}{\bf j}_s({\bf k}),\ \
{\bf A}_s({\bf r})=\sum_{\bf k} e^{i{\bf k}\cdot{\bf r}}{\bf A}_s({\bf k}),
\label{ft}
\end{equation}
where the summation goes over the reciprocal lattice vectors ${\bf k}=(k_x,k_y)
\equiv(2\pi k_1/L_x,2\pi k_2/L_y)$ and $(k_1,k_2)=0,\pm 1,\pm 2,\dots$,
 Eq.(\ref{maxw2}) implies that
\begin{equation}
{\bf A}_s({\bf k}) = {4\pi\over c}{{\bf j}_s({\bf k})\over k^2}, \ \ \
{\bf k}\neq 0.
\label{maxw3}
\end{equation}
Thus, one obtains for the Fourier components of the screening field
\begin{equation}
{\bf h}_s({\bf k}) = {4\pi i\over c}{{\bf k}
\times{\bf j}_s({\bf k})\over k^2}, \ \ \ {\bf k}\neq 0.
\label{hs1}
\end{equation}
In order to evaluate this expression it is helpful to write the supercurrent
(\ref{eq:j}) using wavefunctions $\Psi_\pm$ instead of $s$ and $d$:
\begin{equation}
{\bf j}_s({\bf r})=\delta{e^*\hbar\over 4m}\sum_{\alpha=\pm} \left[
\hat x(1-\alpha\epsilon_v)\Psi_\alpha^*\Pi_x\Psi_\alpha +
\hat y(1+\alpha\epsilon_v)\Psi_\alpha^*\Pi_y\Psi_\alpha + {\rm c.c.} \right].
\label{js1}
\end{equation}
In order to model the layered structure of cuprate superconductors we have
introduced the usual
 geometrical factor $\delta=$(layer thickness/layer spacing).
The case $\delta= 1$
corresponds to the cubic lattice, while $\delta\to
0$ represents the  limit of a single isolated layer.
In this notation, the  Fourier components of  the supercurrent are
\begin{equation}
{\bf j}_s({\bf k})=\delta{e^*\hbar\over 4m}\sum_{\alpha=\pm} \left[
\hat x(1-\alpha\epsilon_v)\langle\Psi_\alpha^*\Pi_x\Psi_\alpha\rangle_{\bf k} +
\hat y(1+\alpha\epsilon_v)\langle\Psi_\alpha^*\Pi_y\Psi_\alpha\rangle_{\bf k}
 + \langle{\rm c.c.}\rangle_{-{\bf k}}  \right].
\label{js2}
\end{equation}
Here we have introduced a shorthand notation
\begin{equation}
\langle\dots\rangle_{\bf k} \equiv {1\over L_x L_y}\int_0^{L_x}dx\int_0^{L_y}dy
\dots e^{-i{\bf k}\cdot{\bf r}},
\label{ift}
\end{equation}
which will prove very convenient in the subsequent calculations.
With some effort, the following useful relations can be derived:
\begin{eqnarray}
\langle\Psi_\alpha^*\Pi_x\Psi_\alpha\rangle_{\bf k} &=& {\sigma_\alpha\over 2}
\left({k_x\over \sigma_\alpha}-ik_y\sigma_\alpha\right)
\langle|\Psi_\alpha|^2\rangle_{\bf k}, \nonumber \\
\langle\Psi_\alpha^*\Pi_y\Psi_\alpha\rangle_{\bf k} &=& {i\over 2\sigma_\alpha}
\left({k_x\over \sigma_\alpha}-ik_y\sigma_\alpha\right)
\langle|\Psi_\alpha|^2\rangle_{\bf k}.
\label{rel}
\end{eqnarray}
In the real space, $\Psi_\alpha({\bf r})$ is a linear combination of
Gaussians, and
thus the Fourier components $\langle|\Psi_\alpha|^2\rangle_{\bf k}$ are easily
evaluated. One obtains
\begin{equation}
\langle|\Psi_\alpha|^2\rangle_{\bf k}={a_{\bf k}\over L_x}
\exp\left\{- { l^2\over 4} [(k_x/\sigma_\alpha)^2 +(k_y\sigma_\alpha)^2]
\right\},
\label{psi2}
\end{equation}
where
\begin{equation}
a_{\bf k}=e^{i{\pi\over 2} k_1k_2}[c_0c^*_{k_2} + (-1)^{k_1}c_1c^*_{k_2+1}]
\label{ak}
\end{equation}
are  real constants, independent of particular lattice shape.
Substituting relations (\ref{rel}) in the expression for the supercurrent
(\ref{js2}) one obtains
\begin{equation}
{\bf j}_s({\bf k})=i\delta{e^*\hbar\over 4m}\sum_{\alpha=\pm} \left[
\hat x(1-\alpha\epsilon_v)(-k_y\sigma^2_\alpha) +
\hat y(1+\alpha\epsilon_v)(k_x/\sigma^2_\alpha)
 \right]\langle|\Psi_\alpha|^2\rangle_{\bf k}.
\label{js3}
\end{equation}
This expression is particularly useful for numerical evaluation of the real
space supercurrent distribution, since in view of the Gaussian dependence of
$\langle|\Psi_\alpha|^2\rangle_{\bf k}$ on ${\bf k}$ [cf. Eq.(\ref{psi2})] it
is
clear that the corresponding Fourier series will converge very rapidly.

Finally, we are in the position to give the local screening field. Substitution
of the above equation (\ref{js3}) into the Maxwell equation (\ref{hs1}) yields
all the Fourier components of the field with ${\bf k}\neq 0$. The ${\bf k}=0$
component is just the real space average of the screening
field $\langle h_s\rangle$ given by
Eq.(\ref{ind}). Combining these results
we obtain, after some algebra, the real space
field distribution of the form
\begin{equation}
{\bf h}_s({\bf r})=-\hat z\pi\delta{e^*\hbar\over mc} \left[
z_0\sum_{\bf k} e^{i{\bf k}\cdot{\bf r}}\langle
|\Psi_+|^2+|\Psi_-|^2\rangle_{\bf k} +
z_1\sum_{{\bf k}\neq 0} e^{i{\bf k}\cdot{\bf r}}\left({k_x^2-k_y^2 \over
k_x^2+k_y^2}\right)
\langle |\Psi_+|^2-|\Psi_-|^2\rangle_{\bf k} \right],
\label{hs2}
\end{equation}
where we have defined numerical factors
\renewcommand{\arraystretch}{1.5}
\begin{equation}
\begin{array}{rl}
z_0 &=[(\sigma_-^2 + \sigma_+^2) +\epsilon_v(\sigma_-^2 - \sigma_+^2)]/2,
\\
z_1 &=[(\sigma_-^2 - \sigma_+^2) +\epsilon_v(\sigma_-^2 + \sigma_+^2)]/2.
\end{array}
\label{zz}
\end{equation}
We notice that the first Fourier sum in the brackets of Eq.(\ref{hs2})
is equal to
$|\Psi_+({\bf r})|^2+|\Psi_-({\bf r})|^2\equiv2(|s({\bf r})|^2 +
|d({\bf r})|^2)$.
Thus in the limit of pure $d$-wave state where $\epsilon_v\to 0$ and
$\sigma_\pm\to 1$
the correspondence with the Abrikosov result for a conventional superconductor
becomes transparent. In this limit we have $z_0\to 1$, $z_1\to 0$ and
$|s({\bf r})|\to 0$, and the spatially varying
 form of the Abrikosov first identity is
recovered, with $d({\bf r})$ playing the role of the conventional order
parameter $\Psi({\bf r})$. The second
sum clearly  has a  nonlocal dependence on the order
parameters and can be written
as $\int d^2r'g({\bf r}-{\bf r}')[s({\bf r}')d^*({\bf r}')+ s^*({\bf r}')
d({\bf r}')]$. Such a term has
no counterpart in the conventional theory, and arises only as a
result of mixing between $s$- and $d$-components of the order parameter. The
nonlocality of this term is a direct consequence of the symmetry of the
problem,
 since by  itself the term  $(sd^*+s^*d)$ is not invariant under $D_4$,
it can enter only in combination with other terms of proper symmetry.


\subsection{Structure of the vortex lattice}

As mentioned above, in order to determine the shape of the vortex lattice,
one needs to evaluate the averages of fourth order terms
$\langle f_4\rangle$ and $\langle h_s^2\rangle$. Now that the distribution
of the magnetic field ${\bf h}({\bf r})$ has been derived, evaluation of these
averages is a straightforward, albeit quite a lengthy procedure. The technical
details of this calculation are worked out in the Appendix, and here we only
summarize the results and discuss some of the physical implications.

Equations (\ref{f44}) and (\ref{hssa}) of the Appendix
give the expressions for the fourth
order averages $\langle f_4\rangle$ and $\langle h_s^2\rangle$ in terms
of rapidly converging sums that are suitable for numerical evaluation.
Making use of these,
the Abrikosov ratio and Ginzburg-Landau parameter
can be expressed in the following simple way
\begin{equation}
\beta_A=
{L_x^2\over 4z_0^2} {\sum_{\bf k}}'
[(z_0+z_1\eta_{\bf k})\Omega_{++}({\bf k})+
(z_0-z_1\eta_{\bf k})\Omega_{--}({\bf k})]^2,
\label{betaA2}
\end{equation}
and
\begin{equation}
\kappa^2=
{4\sum_{\bf k}'
\sum_{\alpha}
[\mu_1\Omega^2_{\alpha\alpha}({\bf k})
+\mu_2\Omega_{\alpha\alpha}({\bf k})\Omega_{\bar\alpha\bar\alpha}
({\bf k})
+\mu_3\Omega_{\alpha\alpha}({\bf k})\Omega_{\alpha\bar\alpha}({\bf k})
+\mu_4\Omega^2_{\alpha\bar\alpha}({\bf k})]
\over
 \pi\delta^2\left({e^*\hbar/ mc}\right)^2
\sum_{\bf k}'
[(z_0+z_1\eta_{\bf k})\Omega_{++}({\bf k})+
(z_0-z_1\eta_{\bf k})\Omega_{--}({\bf k})]^2},
\label{kappa}
\end{equation}
where the prime on the sums means that only  terms with $k_1$ and $k_2$ both
even or odd  are included,
$\Omega_{\alpha\beta}({\bf k})$ are  Gaussian functions
given explicitly
by Eq.(\ref{omegadef}), $\eta_{\bf k}$ is a simple function defined by
Eq.(\ref{eta}, and $\bar{\alpha}=-\alpha$.
Note that the above expression for $\beta_A$ is independent of
parameters $\beta_i$ that enter the quartic part of the free energy $f_4$, and
other quadratic parameters enter only via the
variational parameters $\sigma_\pm$.

The shape of the vortex lattice unit cell is determined
by the ratio $R=L_x/L_y$. The value of $R$ that corresponds to the
thermodynamically  stable configuration, $R_{min}$,
 is obtained by requiring that the Gibbs
free energy  is minimum. Equation (\ref{ggl4}) shows that, at
given external magnetic field $H$, the Gibbs free energy $\langle g \rangle$
is entirely determined by the two parameters given  above, $\beta_A$ and
$\kappa$.
Numerical evaluation of these parameters confirms that $\kappa$
is only very weakly dependent on the particular lattice shape, as it is
illustrated  by Fig.\ \ref{fig:beta}. The dependence
of the Gibbs free energy (\ref{ggl4}) on $R$ is almost entirely contained in
the Abrikosov ratio $\beta_A$ and thus, in most of the parameter
space, the
minimum of $\langle g\rangle$ coincides to a good accuracy with the minimum
of $\beta_A$. For example in the particular case displayed in the Fig.\
\ref{fig:beta} the minimum of $\beta_A$ differs by less than 2\% from the
minimum the full free energy.

Fig.\ \ref{fig:beta} also shows a typical dependence of $\beta_A$ on $R$ for
different values of mixed gradient coupling $\epsilon_v$, as obtained by
numerical  evaluation of Eq.(\ref{betaA2}).  When $\epsilon_v=0$ the
superconductor is in a pure $d$-wave state with no $s$-wave component
present. Within the phenomenological GL theory, this situation is identical
to the case of a conventional superconductor studied by Abrikosov. Thus,
the state with minimum free energy has $R_{min}=\sqrt{3}$
which corresponds to the
usual triangular vortex lattice. In this limit we obtain the correct
value of $\beta_A=1.1596$ as quoted by Kleiner {\em et al.}\cite{kleiner}
However, as soon as non-zero coupling $\epsilon_v$ is introduced, the
situation changes
and the minimum of $\beta_A$ shifts to the values  $R_{min}<\sqrt{3}$,
signalling that an oblique vortex lattice is favored. The minimum $R_{min}$
varies continuously with $\epsilon_v$ and at certain value of $\epsilon_v$,
which depends
on other parameters in GL free energy, $R_{min}$ reaches the value of 1,
corresponding to the square lattice. Further increase of $\epsilon_v$ then
has no effect on the shape of the lattice, which remains square.

One may conclude that in a $d$-wave superconductor, in the regime close to the
upper critical field $H_{c2}$, a general oblique vortex lattice is
thermodynamically stable, unless the material is in one of the limiting
regimes in which the
mixed gradient coupling $\epsilon_v$ is very small or very large.
Numerical \cite{soininen} and analytical \cite{ren} calculations based on the
simple mean field model with proper symmetries, find evidence for a
mixed gradient term of about the same order of magnitude as the
conventional gradient terms. This would seem to argue against the two
limiting cases mentioned above.

An example of the oblique vortex lattice is
displayed in Fig.\ \ref{fig:dscont}, where we show the $d$ and $s$ components
of the order parameter as obtained by numerical evaluation of Eqs.(\ref{drsr}),
for given set of GL parameters. An interesting conclusion can be drawn by
comparing the two components of the order parameter: it is evident that the
non-trivial nodal structure of the $s$-wave component, such as  was described
in
Sec.\ III for an isolated vortex, persists in this high field regime. Indeed,
zeros of $s$ are present in the regions where $|d|>0$. This  quite
remarkable result appears to suggest that the ``extra'' vortices in
the $s$-component are present over the entire portion of the phase diagram
representing the mixed state of a $d$-wave superconductor.

Many experimental probes are sensitive to the spatial variations of the
magnetic field rather than to the order parameter itself.
The spatially varying component of the magnetic field, $h_s({\bf r})$, as
evaluated from  Eq.(\ref{hs2}) is shown in Fig.\ \ref{fig:hcont}. Notice that
as a consequence of the Maxwell equation
$\nabla\times{\bf h}_s=(4\pi/c){\bf j}_s$,
it follows that the contours of constant magnetic field coincide with the
supercurrent streamlines. Comparison to the order parameter plot in
Fig.\ \ref{fig:dscont} confirms that magnetic field and supercurrent
distributions have the same symmetry as the vortex lattice. A non-trivial nodal
structure of the $s$-wave has an effect on the field distribution, which
develops two nonequivalent saddle points, marked S1 and S2 in
Fig.\ \ref{fig:hcont}. In principle, it might be possible to determine
such structure by $\mu$SR or NMR experiments. Fig.\ \ref{fig:musr} displays the
$\mu$SR line shapes that result from the magnetic field distribution as
discussed above. The quantity shown is
\begin{equation}
P(h)={1\over L_xL_y}\int\delta[h-h({\bf r})]d^2r,
\end{equation}
for the case of triangular, oblique and square flux lattices. In the triangular
and square lattices symmetry requires only one type of saddle point, resulting
in the conventional single peak structure. In the oblique lattice, which is
characteristic of a $d$-wave superconductor, the two non-equivalent saddle
points give rise to two distinct Van Hove type singularities.
Appearance of two distinct peaks in $\mu$SR or NMR spectra,
 would provide  evidence for $d$-wave behavior, since the
explanations of oblique vortex lattice that invoke anisotropy within single
component model\cite{walker} do not lead to this effect.

The last subject that we want to address here concerns the spatial
orientation of the vortex lattice with respect to the crystalline axes of
the superconductor.
{}From  Fig.\ \ref{fig:dscont}, it can be seen that the principal axes of
the vortex lattice are not aligned with any of the high symmetry directions
of the underlying crystal. Instead, it is
the (110) direction of the vortex lattice that
coincides with the (100) or (010) directions of the crystal. It turns out that
 the construction of the vortex lattice
as presented above forces this particular
orientation and does not allow for  identical configurations that are
rotated by some angle $\alpha$. In the traditional one component case, this
is not a concern since the free energy has full rotational invariance.
In the present case, however, we must take a closer look at these rotated
configurations as we have terms in the free energy that break rotational
invariance. It is conceivable that such rotated configurations might in fact be
lower in free energy than the ones we have considered so far. In what follows
we show by an explicit calculation that this is not the case, and that we have
in fact found the solution that corresponds to the absolute
minimum of $f$ as given by Eq.(\ref{fgl}).

Consider a simple rotation of the coordinate system in the $xy$ plane by
an angle $\alpha$
\begin{eqnarray}
x &=& x'\cos\alpha - y'\sin\alpha, \nonumber \\
y &=& x'\sin\alpha + y'\cos\alpha. \label{rot}
\end{eqnarray}
Under such transformation all the second order terms are invariant except
for the mixed gradient term which transforms as follows
\begin{eqnarray}
{\partial s \over \partial y} {\partial d^* \over \partial y} -
{\partial s \over \partial x} {\partial d^* \over \partial x} + {\rm c.c.} &=&
(\cos^2\alpha-\sin^2\alpha)\biggl(
{\partial s \over \partial y'} {\partial d^* \over \partial y'} -
{\partial s \over \partial x'} {\partial d^* \over \partial x'}
\biggr) \nonumber \\&&  \ \ \ \ \ +
2\sin\alpha\cos\alpha \biggl(
{\partial s \over \partial x'} {\partial d^* \over \partial y'} +
{\partial s \over \partial y'} {\partial d^* \over \partial x'}
\biggr) + {\rm c.c.}. \label{mixed}
\end{eqnarray}
One can now derive and analyze the linearized field equations using the rotated
coordinates $(x',y')$ in  exactly the same way as before,
 and the angle $\alpha$
becomes just another variational parameter with respect to which
the free energy is minimized. It turns out that it is possible to write down
the linearized equations for $s$ and $d$ that are
identical to Eqs.(\ref{eq:sdlin2})
with $V$ changed to  $V=\epsilon_v(\hbar\omega_c/2)(e^{2i\alpha} a^\dagger
a^\dagger +
e^{-2i\alpha}aa)$. In such case, one expects there will be a constant phase
difference $2\alpha$ between $s$ and $d$ components, and the appropriate
variational solution is of the form
\begin{equation}
d(x) = e^{-i\alpha}[{\varphi^+}(x)+{\varphi^-}(x)], \ \
s(x) = e^{i\alpha}[{\varphi^+}(x)-{\varphi^-}(x)],
\label{rotvar}
\end{equation}
where ${\varphi^+}$ and ${\varphi^-}$ are the normalized lowest
eigenfunctions of
harmonic oscillator as defined by Eq.(\ref{varsol}). The energy eigenvalue
is easily evaluated, and we obtain a generalization of  Eq.(\ref{Emin3})
\begin{eqnarray}
{\langle E\rangle \over{\Delta T}} = {T - T^*\over{\Delta T}}
&+& {1\over 4}\biggl({\hbar\omega_c\over{\Delta T}}\biggr)
\biggl[(1+\epsilon_v\cos^2 2\alpha)\tan\vartheta
+(1-\epsilon_v\cos^2 2\alpha){1\over\tan\vartheta}\biggr] \nonumber\\
&-& {1\over 2}\sqrt{{2\tan\vartheta \over 1+\tan^2\vartheta}}
\biggl[ 1+\epsilon_v\sin^2 2\alpha \biggl({\hbar\omega_c\over{\Delta T}}\biggr)
{1-\tan^2\vartheta \over 1 + \tan^2\vartheta} \biggr].\label{Eminrot}
\end{eqnarray}
It is a simple matter to minimize $\langle E \rangle$ with respect
to $\alpha$, and one finds that minima can occur only for $\alpha=0,\pm\pi/2,
\pm\pi$. Thus, we are led to the conclusion that within our
variational solution the most stable vortex lattice is the one aligned
with the underlying crystal as described above (cf. Fig.\ \ref{fig:dscont}).
Among the fourth order terms in the free energy only $\beta_4(s^{*2}d^2 +
s^2d^{*2})$ depends on $\alpha$. This dependence is particularly simple;
upon rotation the constant $\beta_4$ changes to  $\beta_4\cos4\alpha$.
Clearly, this term only has minima for trivial values of $\alpha=0,\pm\pi/2,
\dots$, so the above conclusion should hold even when the fourth order terms
are included. In order to verify that this conclusion is not altered by some
complicated interplay between angular dependencies of $f_2$ and $f_4$, we have
carried out the numerical minimization of the free energy of the rotated
vortex lattice, along the lines indicated for the case $\alpha=0$. We
find that, for all the regions of the parameter space that were investigated,
the absolute minimum of the free energy occurs for $\alpha=0$. As a
consistency check we have also
verified that identical minima are found for $\alpha=
\pm\pi/2,\pm\pi$, which corresponds to the discrete rotations under the
$D_4$ group.

The above conclusion concerning the  orientation of the vortex
lattice may be understood by analyzing the mixed gradient term in the free
energy density (\ref{fgl}). Its structure forces the vortex lattice to align
in such a way that the greatest gradient of order parameters is along one
axis, while the smallest possible gradient is along other axis. An arrangement
of vortices such as the one  shown in Figure \ref{fig:dscont} definitely
satisfies this requirement.


\section{Summary and discussion}

The main goal of this work was to present a detailed study of the vortex state
in a $d$-wave superconductor, focusing on the properties arising from $s$-$d$
mixing that have no analogue in conventional superconductors. Analysis of the
vortex state is done in two regimes; in the vicinity of $H_{c1}$ where the
properties of isolated vortex lines can be studied, and near $H_{c2}$ where
the collective properties of vortices forming a dense lattice are important.

For the single vortex line the topological structure of the induced $s$-wave
order parameter is highly non-trivial, consisting of one counter-rotating
unit vortex, centered at the core, surrounded by four additional positive
vortices located symmetrically at a distance of several coherence lengths from
the core. A new result of this work is the realization that the above structure
will occur for all parameter configurations that give rise to stable
$d$-wave in the bulk (provided one is well into the type-II regime),
and not only in the vicinity of $T_c$ as was originally
suggested.\cite{berlinsky} This conclusion is confirmed by an explicit
integration  of the GL equations over the wide range of parameters, and also by
the calculations of Ichioka {\em et al.}\cite{ichioka}, who find analogous
behavior using the quasi-classical Eilenberger equations. The
question arises as to
whether this non-trivial topological structure of a single vortex could be
probed
experimentally. There are clearly many complicating factors which are likely
to render this task very difficult. The main challenge arises from the fact
that
one expects the induced $s$-component to be small, on the order of few percent
of $d$. Such a small admixture of $s$ might be hard to detect directly, and
the corresponding distortion of the $d$-wave amplitude, supercurrent and
magnetic field distributions will also be small. It might in principle be
possible to probe the $s$-component by scanning Josephson tunneling from an
$s$-wave tip, which by symmetry would not couple to the dominant $d$-wave.
With sufficient resolution such an
experiment could detect strong anisotropy in the
$s$-component and possibly also the extra nodes. The internal structure of a
vortex will also have an effect on the transport properties; e.g. it is
conceivable that it may lead to  changes in the Magnus
 force acting on a vortex in a current field. These issues clearly require
further investigation.

Finally we note that although a finite induced
$s$-component will restore the gap along the $|k_x|=|k_y|$ directions in the
vicinity of the core, this will not invalidate the prediction of
Volovik\cite{volovik} regarding the $\sim \sqrt{H}$ contribution to the density
of states (DOS) on the Fermi surface, which was recently confirmed by specific
heat measurement by Moler {\em et al.}\cite{moler} Volovik's prediction is
based
on the observation (originally used by Yip and Sauls\cite{yip} to predict the
nonlinear Meissner effect) that the quasiparticle excitation spectrum is
shifted by the
superfluid velocity field around the vortex core, with the dominant
contribution
coming from quasiparticles far from the core in position space and close
to the nodes in momentum space. Since the amplitude of the $s$-component far
from the core vanishes as $1/r^2$ the reduction of the
DOS will be always negligible
beyond a certain distance from the core compared to the energy shift due to
superfluid velocity which decays only as $1/r$. Thus at
relatively small fields compared to $H_{c2}$, such as were used in the specific
heat measurements,\cite{moler} there
will be no correction to the Volovik's result from the
induced $s$-wave. At stronger fields, when the vortices are closely spaced,
corrections may appear; however, in such a case one expects Volovik's
derivation to break down since the concept of an isolated vortex with a well
defined asymptotic flow field loses its meaning in the dense Abrikosov
lattice.

The vortex lattice near $H_{c2}$
is in general oblique for a $d$-wave superconductor.
The precise shape determined by an angle $\phi$ between primitive vectors
depends in a complicated way on the parameters in the
GL free energy; most strongly on the
mixed gradient coupling $\epsilon_v$ and on magnetic field via the parameter
$\Lambda=\hbar\omega_c/\Delta T$, which also determine the relative magnitude
of $s$. Quite generally, when $\epsilon_v=0$, the $s$-component vanishes and
the lattice
is triangular. By increasing $\epsilon_v$ and $\Lambda$ the lattice is
continuously
deformed, becoming oblique and eventually square. Observation of
an oblique flux lattice with $\phi\simeq 73^\circ$ was reported by Keimer
{\em et al.}\cite{keimer1} using small angle neutron scattering from YBCO in
magnetic fields 0.5 T$\leq H\leq$ 5 T. This would be in agreement with our
result, although as was pointed out by Walker and Timusk,\cite{walker} the
observed distortion may also be accounted for by the intrinsic $a$-$b$ plane
anisotropy of the orthorhombic
YBCO crystal. More recently an oblique vortex lattice
with $\phi\simeq 77^\circ$ was found in YBCO using STM by Maggio-Aprile
{\em et al.}\cite{renner} This technique also revealed elongated vortex cores
with the ratio of principal axes about 1.5. If, as noted by authors, this
elongation was due to the $a$-$b$ anisotropy in coherence lengths, within a
simple London model of $s$-wave superconductivity this would lead to the flux
lattice with an angle inconsistent with the actual observed value of
$77^\circ$.
Thus it would appear that the $a$-$b$ anisotropy alone cannot explain the
observed distortion in the vortex lattice and additional effects, such as the
internal symmetry of the order parameter, must be invoked in order to account
for the experimental data. In this respect it would be most interesting to see
if an oblique lattice can also be established experimentally in truly
tetragonal
superconductors. Alternatively it would be desirable to study the analogous
GL theory for the $D_2$ orthorhombic symmetry; unfortunately such a theory is
complicated and contains even more phenomenological parameters so that
a quantitative comparison with experiment would be difficult.\cite{ortho}
An alternative way of distinguishing between the effects of $a$-$b$ anisotropy
and $d$-wave symmetry is to study the magnetic field distributions in the
vortex
lattice. The present theory predicts a double-peak structure in $\mu$SR or
NMR lineshapes whenever the flux lattice is oblique, while interpretations
based on simple scaling arguments\cite{walker} lead to conventional single-peak
lineshapes.

\acknowledgements

The authors are grateful to  D. L. Feder, R. Joynt, M. Kvale, K. A. Moler,
J. A. Sauls,  M. Sigrist, and S. K. Yip for valuable discussions on the
subject.
This work has been partially supported by the Natural Sciences and
Engineering Research Council of Canada, Ontario Centre for Materials
Research, and by the National Science Foundation under Grant
No. DMR-91-20361. M. F.   gratefully acknowledges support from a NATO
Science Fellowship. We would also like to thank the Aspen Center for
Physics, where this collaboration was initiated and part of the work was done.

\appendix
\section*{Evaluation of quartic averages \lowercase{$\langle f_4\rangle$} and
\lowercase{$\langle h_s^2\rangle$}}
We first evaluate  the contribution of  $\langle f_4 \rangle$.
For the purposes of calculation
it is convenient to express $\langle f_4 \rangle$
 in terms of the functions $\Psi_\pm$,
\begin{equation}
\langle f_4 \rangle=
\mu_1\langle |\Psi_+|^4\rangle +
\mu_2\langle|\Psi_+|^2|\Psi_-|^2\rangle +
\mu_3\langle|\Psi_+|^2 \Psi_+\Psi_-^*\rangle  +
\mu_4\langle\Psi_+^2\Psi_-^{*2}\rangle
+\Bigl[ \Psi_+\leftrightarrow \Psi_- \Bigr],
\label{f4}
\end{equation}
where the constants $\mu$ are given as follows
\begin{equation}
\begin{array}{rccccl}
\mu_1={1\over 16}(&\beta_1& +\beta_2& +\beta_3 &+2\beta_4&), \\
\mu_2={2\over 16}(&\beta_1& +\beta_2&          &-2\beta_4&), \\
\mu_3={4\over 16}(&-\beta_1&+\beta_2&          &         &), \\
\mu_4={1\over 16}(&\beta_1& +\beta_2& -\beta_3 &+2\beta_4&).
\end{array}
\label{mu}
\end{equation}
The easiest way to evaluate the spatial averages is to express them
as  Fourier series. For example, one can write the typical member as
follows,
\begin{equation}
\langle\Psi_+^{*2}\Psi_-^2\rangle =\sum_{\bf k} \langle\Psi^*_+\Psi_-
\rangle_{\bf k}\langle\Psi^*_+\Psi_-\rangle_{-{\bf k}},
\label{dec}
\end{equation}
where we have used only the basic properties of the Fourier series. The utility
of this formulation lies in the fact that components of the form
$\langle\Psi^*_\alpha\Psi_\beta\rangle_{\bf k}$ can be expressed in terms of
simple Gaussians, and consequently the summations indicated in Eq.(\ref{dec})
converge very rapidly. In particular it is useful to define
\begin{equation}
\langle\Psi^*_\alpha\Psi_\beta\rangle_{\bf k} = a_{\bf k}
\Omega_{\alpha\beta}({\bf k}),
\label{omega}
\end{equation}
where the coefficients $a_{\bf k}$ are given by Eq.(\ref{ak}). The factors
$\Omega_{\alpha\beta}({\bf k})$ contain all the dependence on the lattice
structure and can be evaluated by explicit integration; we have
\begin{equation}
\begin{array}{rl}
\Omega_{\alpha\alpha}({\bf k})&={1\over L_x}
\exp\left\{-{ l^2\over 4} [(k_x/\sigma_\alpha)^2
+(k_y\sigma_\alpha)^2]\right\},
\\
\Omega_{\alpha\beta}({\bf k})&={1\over L_x}
\sqrt{2\over \sigma_+^2 + \sigma_-^2}
\exp\left\{-{ l^2\over 4} {2\over \sigma_+^2 + \sigma_-^2}
[k_x^2 +k_y^2 + i(\sigma_\alpha^2-\sigma_\beta^2)k_xk_y]\right\}, \ \
\alpha\neq\beta.
\end{array}
\label{omegadef}
\end{equation}
In terms of these functions, $\langle f_4 \rangle$ can be written in a
 compact form,
\begin{equation}
\langle f_4 \rangle =\sum_{\bf k} a_{\bf k} a_{-{\bf k}} \sum_{\alpha=\pm}
[\mu_1\Omega^2_{\alpha\alpha}({\bf k})
+\mu_2\Omega_{\alpha\alpha}({\bf k})\Omega_{\bar\alpha\bar\alpha}({\bf k})
+\mu_3\Omega_{\alpha\alpha}({\bf k})\Omega_{\alpha\bar\alpha}({\bf k})
+\mu_4\Omega^2_{\alpha\bar\alpha}({\bf k})],
\label{f44}
\end{equation}
which is suitable for numerical evaluation. In deriving Eq.(\ref{f44}) we have
used the symmetry
$\Omega_{\alpha\beta}({\bf k})=\Omega_{\alpha\beta}(-{\bf k})$
which is apparent from Eqs.(\ref{omegadef}), and we use the notation
$\bar\alpha=-\alpha$.

Let us now turn to calculation of $\langle h_s^2\rangle$. A similar approach as
above will work here if we write
\begin{equation}
\langle h_s^2 \rangle = \sum_{\bf k}  h_s({\bf k}) h_s(-{\bf k}).
\label{hsa}
\end{equation}
The Fourier components $h_s({\bf k})$ can be deduced from Eq.(\ref{hs2}),
\begin{equation}
h_s({\bf k})=-\pi\delta{e^*\hbar\over mc} \left[
z_0\langle |\Psi_+|^2+|\Psi_-|^2\rangle_{\bf k} +
z_1\eta_{\bf k}\langle |\Psi_+|^2-|\Psi_-|^2\rangle_{\bf k} \right],
\label{hs2k}
\end{equation}
where we have introduced the quantity
\begin{equation}
\eta_{\bf k}=\left\{
\begin{array}{ll}
 {{k_x^2-k_y^2\over k_x^2+k_y^2}},\ & {\rm if}\ {\bf k}\neq 0 \\
 0,                               & {\rm if}\ {\bf k}= 0
\end{array}
\right.
\label{eta}
\end{equation}
which allows  all of the Fourier components
of $h_s({\bf k})$ to be expressed by a single equation (\ref{hs2k}).
A compact  expression for the average squared field can be
written in terms of the functions $\Omega_{\alpha\beta}({\bf k})$ and has the
 following form
\begin{equation}
\langle h_s^2 \rangle = \pi^2\delta^2\left({e^*\hbar\over mc}\right)^2
\sum_{\bf k}  a_{\bf k} a_{-{\bf k}}
[(z_0+z_1\eta_{\bf k})\Omega_{++}({\bf k})+
(z_0-z_1\eta_{\bf k})\Omega_{--}({\bf k})]^2.
\label{hsa2}
\end{equation}
Let us finally write, for the purpose of completeness, the expression for the
average field:
\begin{equation}
\langle h_s\rangle \equiv h_s({\bf k}=0)=-\pi\delta
 z_0 a_0\left({ e^*\hbar\over mc}\right)
[\Omega_{++}(0)+\Omega_{--}(0)]
=-{2\pi\delta}{z_0\over L_x}
 \left({ e^*\hbar\over mc}\right) (|c_0|^2 + |c_1|^2).
\label{hssa}
\end{equation}

We have now expressed all the averages that are
needed to minimize the Gibbs free energy (\ref{ggl4}). In principle, both
$\beta_A$ and $\kappa$ can now be evaluated numerically using the rapidly
converging sums (\ref{f44}) and (\ref{hsa2}). However, following the work
of Kleiner, Roth and Autler, \cite{kleiner} it is practical to carry out
the  minimization with respect to the constants $c_0$ and $c_1$ analytically.
This reduces the relevant parameter space to a single parameter, the geometric
ratio $R=L_x/L_y$. To this end we have purposely singled out the  dependence
on these parameters in the expressions for $\langle f_4 \rangle$ and
$\langle h_s^2 \rangle$. In particular, the entire dependence on $c_0$ and
$c_1$
is contained in the constants $a_{\bf k}$, and both  $\langle f_4 \rangle$ and
$\langle h_s^2 \rangle$ are of the form
\begin{equation}
\sum_{{\bf k}}a_{\bf k} a_{-{\bf k}} f(k_1,k_2) =
\sum_{k_1k_2}
\left\{e^{i{\pi\over 2} k_1k_2}[c_0c^*_{k_2} + (-1)^{k_1}c_1c^*_{k_2+1}]
\right\}^2 f(k_1,k_2),
\label{sumak1}
\end{equation}
where $f(k_1,k_2)$ is independent on  $c_0$ and $c_1$.
Our goal here is to factor out the entire dependence of this expression on
the constants $c_0$ and $c_1$. This is done by considering separately
the cases when $k_1$, $k_2$ are even and odd, and recalling that by
assumption $c_{2k}=c_0$ and $c_{2k+1}=c_1$.
We obtain an expression of the following form
\begin{eqnarray}
&&(|c_0|^4+|c_1|^4)\sum_{k_1k_2}[f(2k_1,2k_2) + f(2k_1+1,2k_2)]  \nonumber \\
&+&
(c_0^2c_1^{*2}+c_0^{*2}c_1^2)\sum_{k_1k_2}[f(2k_1,2k_2+1) - f(2k_1+1,2k_2+1)]
\\
&+&2|c_0|^2|c_1|^2\sum_{k_1k_2}[f(2k_1,2k_2) - f(2k_1+1,2k_2) +
                             f(2k_1,2k_2+1) + f(2k_1+1,2k_2+1].\nonumber
\end{eqnarray}
Clearly, the entire denominator
$8\pi \langle f_4 \rangle - \langle h_s^2 \rangle$ of the free energy
(\ref{ggl3}) can be written in the above form, where  the dependence
on constants $c_0$ and $c_1$ is explicitly shown. Combining this with the
expression (\ref{hssa}) for the average induced field $ \langle h_s\rangle$
it is easy to see that the total free energy (\ref{ggl3}) can be written
schematically as
\begin{equation}
 -{(|c_0|^2+|c_1|^2)^2 \over
(|c_0|^4+|c_1|^4)G_0(R) + 2|c_0|^2|c_1|^2G_1(R) +
2{\rm Re}(c_0^2c_1^{*2})G_2(R) },
\label{cmin1}
\end{equation}
where $G_i(R)$ are complicated functions of $R$ and other GL parameters, but
are independent of $c_0$ and $c_1$. The above expression can be easily
minimized with respect to $c_0$ and $c_1$; one obtains that a condition for
the minimum is $c_1=\pm ic_0$.
Clearly, the value of the expression (\ref{cmin1})
only depends on the ratio $c_1/c_0$, so we can arbitrarily choose
\begin{equation}
c_0=1,\ \ c_1=i.
\label{cmin2}
\end{equation}
With this choice, we have an identity
\begin{equation}
a_{\bf k} a_{-{\bf k}} = [(-1)^{k_1} + (-1)^{k_2}]^2,
\label{cmin3}
\end{equation}
which will simplify evaluation of the sums in $\langle f_4 \rangle$ and
$\langle h_s^2 \rangle$. Also, the expression for the average induced field
(\ref{hssa}) simplifies:
\begin{equation}
\langle h_s\rangle
=-4\pi\delta {z_0\over L_x}\left({ e^*\hbar\over mc}\right).
\label{hssa2}
\end{equation}
%


\begin{figure}
\caption[]{Schematic diagram of phases $\theta_d$ and $\theta_s$ of the two
components of the order parameter in the two asymptotic regions close to and
far from the vortex core. Note that this diagram is more complete than the
similar one published in Ref.\ [\onlinecite{soininen}] in that it includes
the region outside the core. The present diagram also differs from that in
Ref.\ [\onlinecite{ren}] which shows (we believe incorrectly) the $s$-component
with opposite overall sign outside the core.}
\label{fig:ph1}
\end{figure}

\begin{figure}
\caption[]{Contour plot of the amplitudes of the (a) $d$-wave and (b) $s$-wave
components of the order parameter. The GL parameters used for the plot are:
$\gamma_s=\gamma_d=\gamma_v$, $\alpha_s=10|\alpha_d|$, $\beta_1=\beta_3=0$,
and $\beta_4=0.5\beta_2$.  The lightest regions of the diagram
correspond to the largest amplitudes. The scale is in units of the $d$-wave
coherence length $\xi_d$.}
\label{fig:Samp}
\end{figure}

\begin{figure}
\caption[]{Amplitude of the $d$-wave and the $s$-wave component
 along the $x$-axis (solid line) and along the diagonal $x=y$ (dotted line)
normalized to the bulk value $d_0$. The parameters used are the same as in
Fig.\ \ref{fig:Samp}. The $d$-component is almost completely isotropic for
this case so that the two cuts are indistinguishable.}
\label{fig:Scut}
\end{figure}

\begin{figure}
\caption[]{The angle of arrow with respect to the horizontal $x$-axis
represents
the phase of the (a) $d$-wave and (b) $s$-wave components
of the order parameter. The parameters used are the same as in
Fig.\ \ref{fig:Samp}.}
\label{fig:Sphsd}
\end{figure}

\begin{figure}
\caption[]{Relative phase $\Delta\theta$ of the two components of the order
parameter, for the same parameters as in Fig.\ \ref{fig:Samp}.}
\label{fig:Sphrel}
\end{figure}

\begin{figure}
\caption[]{Contour plot of (a) supercurrent amplitude (b) supercurrent
streamlines (which coincide with the lines of constant magnetic field), for the
 same parameters as in Fig.\ \ref{fig:Samp}.}
\label{fig:Sj}
\end{figure}

\begin{figure}
\caption[]{Contour plot of the amplitudes of (a) $d$-wave and (b) $s$-wave
components of the order parameter for a different set of GL parameters:
$\gamma_s=\gamma_d=\gamma_v$, $\alpha_s=1.4|\alpha_d|$,
$\beta_1=\beta_2=\beta_3=\beta_4$.}
\label{fig:Samp2}
\end{figure}

\begin{figure}
\caption[]{Contour plot of (a) supercurrent amplitude (b) supercurrent
streamlines (which coincide with the lines of constant magnetic field), for the
 same parameters as in Fig.\ \ref{fig:Samp2}.}
\label{fig:Sj2}
\end{figure}

\begin{figure}
\caption[]{Dependence of upper critical field $H_{c2}(T)$
on temperature for various values of parameter $\epsilon_v$ and
  $T_s=0.5 T_d$, $T=0.75 T_d$.}
\label{fig:hc2}
\end{figure}

\begin{figure}
\caption[]{Abrikosov ratio $\beta_A$ as a function of the lattice geometry
factor $R=L_x/L_y$ for different values of $\epsilon_v$. Note that the minimum
of $\beta_A$
is moving from $R=\sqrt{3}$ to $R=1$ as $\epsilon_v$ increases. This implies a
 continuous
deformation of the initially triangular vortex lattice into an oblique and
square lattice. The parameters used are $T_s=0.5 T_d$, $T=0.75 T_d$,
$\beta_1=\beta_2=\beta_3=\beta_4=1$, and $B=0.8 H_{c2}$. The inset shows the
$R$-dependence of the squared Ginzburg-Landau ratio $\kappa^2$ on
approximately thesame scale.}
\label{fig:beta}
\end{figure}

\begin{figure}
\caption[]{Contour plot of the amplitudes of (a) $d$-component and (b)
$s$-component of the order parameter in the vortex lattice.
The same parameters are used as in
Fig.\ \ref{fig:beta}  with $\epsilon_v=0.45$ resulting in an oblique vortex
lattice with $R_{min}=1.29$ and the angle between primitive vectors
$\phi=76^\circ$. The oblique unit cell containing one flux quantum is marked
by a solid line.}
\label{fig:dscont}
\end{figure}

\begin{figure}
\caption[]{
Distribution of the magnetic field $h_s$ in the vortex lattice. Letters $M$,
$m$, $S1$ and $S2$ denote the
 maximum, minimum and two saddle points respectively.
GL parameters used for the plot are same as in Fig.\ \ref{fig:dscont}.}
\label{fig:hcont}
\end{figure}

\begin{figure}
\caption[]{Typical $\mu$SR lineshapes as obtained from the magnetic field
distribution in the vortex lattice. The curves shown are for triangular
($\epsilon_v=0.0$), oblique ($\epsilon_v=0.4$) and square ($\epsilon_v=0.6$)
flux lattices. Magnetic
field on the horizontal axis is in the units of $h_0=|\langle h_s \rangle |
\equiv 4\pi\delta(z_0/L_x)(e^*\hbar/mc)$ and the curves are offset verticaly
for clarity.}
\label{fig:musr}
\end{figure}


\begin{thebibliography}{99}
\bibitem{wind} The winding number (or vorticity) is defined in the conventional
way\cite{tinkham} as $n_v={1\over 2\pi}\oint_C\nabla\vartheta\cdot d{\bf l}$,
where $\vartheta$ is the phase of the order parameter and $C$ is an arbitrary
closed contour containing the vortex.
\bibitem{footnote}Note that within this scheme
one should in principle add one more variational parameter $\alpha$ by allowing
${\varphi^+}$ and ${\varphi^-}$ to enter the definition of $s$ and $d$ in
Eq.(\ref{newar})
with different weights, $i.e.$, $d={\varphi^+} \cos\alpha +
{\varphi^-}\sin\alpha$, etc. A
careful analysis of this case shows that $E$ is always minimized by
$\alpha=\pi/4$, thus recovering  Eq.(\ref{newar}).
\bibitem{ortho} We have verified by an explicit calculation that adding
second order invariants appropriate for the orthorhombic symmetry to the
free energy does not qualitatively alter our conclusions regarding the vortex
lattice.
\end{thebibliography}
\end{document}